# Multi-level emission impacts of electrification and coal pathways in China's netzero transition

Chen Chris Gong*[1], Falko Ueckerdt[1], Christoph Bertram[1,2], Yuxin Yin[1,3,4], David Bantje[1], Robert Pietzcker[1], Johanna Hoppe[1], Robin Hasse[1], Michaja Pehl[1], Simón Moreno-Leiva[1], Jakob Duerrwaechter[1], Jarusch Muessel[1], Gunnar Luderer[1,5]

[1] Potsdam Institute for Climate Impact Research (PIK), Potsdam, Germany

[2] Center for Global Sustainability, School of Public Policy, University of Maryland, College Park, USA

[3] Laboratory of Urban and Environmental Systems (URBES), Institute of Architecture (IA), School of Architecture, Civil and Environmental Engineering (ENAC), Ecole Polytechnique Fédérale de Lausanne (EPFL), Lausanne, Switzerland

[4] Department of Urban Water Management, Swiss Federal Institute of Aquatic Science and Technology, Dübendorf, Switzerland

[5] Global Energy Systems Analysis, Technische Universität Berlin, Berlin, Germany

## Abstract

Decarbonizing China's energy system requires both greening the power supply and electrifying end-use sectors. However, concerns exist that electrification may increase emissions while coal power dominates. Using a global climate model, we explore electrification scenarios with varying coal phase-out timelines and assess their climate impact on China's sectors. A ten-year delay in coal phase-out could increase global peak temperature by about 0.02°C. However, on a sectoral level, there is no evidence of significant additional emissions from electrification, even with a slower coal phase-out. This challenges the sequential "order of abatement" view, showing electrification can start before the power sector is fully decarbonized. As long as power emission intensity drops below 150 $gCO_2$/kWh by 2040, electrification can substantially reduce the carbon footprint of buildings, steel, and transport services, and along with energy efficiency measures, it can avoid approximately 0.035°C of additional global warming by 2060.

## Context&Scale

Decarbonizing the energy system requires both cleaner power generation and electrifying energy demand. However, some uses of energy are more efficient at turning power into products or services. For less efficient uses, electrification could lead to more $CO_2$ emissions if the power supply isn't green enough. So, when is the best time to electrify, and which types of technologies should be used? In a power mix with 60% coal, like China's, could electrification lead to more emissions? What are the climate impacts of power and energy use? Our study shows that for every 10-year delay in phasing out coal power in China, peak warming could increase by 0.02°C. However, rapid electrification doesn't cause large extra emissions. When combined with green hydrogen and energy efficiency, it can even help avoid an additional 0.035°C of warming.

[1] Corresponding author: chen.gong@pik-potsdam.de

# Introduction

Decarbonizing the energy system depends on green power generation and the transformation of energy demand through two primary approaches: direct and indirect electrification. Direct electrification involves using green electricity directly, as seen in electric vehicles and heat pumps. In contrast, indirect electrification generally converts electricity into hydrogen and other derived molecules. These are then used to replace applications where direct electrification may not be practical or cost-effective, such as in high-temperature industrial processes, long-distance transport (e.g., ships and airplanes), and non-energy use in the chemical industries.

For the last four years, the increase in gasoline price, along with long-standing industrial policies, has accelerated the rapid adoption of battery electric vehicles (BEVs) in China, making the direct electrification of road transport an economic reality[1–3]. The combined annual sales share of hybrid and plug-in electric vehicles rose to above 50% in September 2024[4]. For heavy freight trucks, the first half year figure was as high as 9.21%[5]. Besides the climate imperatives, road transport electrification is aligned with several political and economic considerations, including reducing air pollution from road traffic[6,7], industrial upgrades and continued manufacturing export-led growth especially to Europe[8], and reducing oil import dependencies[9–11].

However, when it comes to the phase-out of coal power – one of the highest emitting sources in China, there are factors that complicate near-term reform towards renewable generation. First, due to strong domestic demand growth, there is growing concern over energy supply adequacy. China's annual electric demand increases rapidly in recent years (6.7% in 2023, 9.8% in Q1 2024), due to the electrification of fishery and livestock sector, a growing energy-intensive digital sector (e.g. artificial intelligence and 5G), as well as air condition use due to heat waves[12,13]. While electrification via renewables provides a way to hedge the risk of a potential oil import embargo in the situation of conflict[9–11], China still has an abundance of coal, which can supply electricity as well as synthetic liquid fuel.

Second, the current regulatory and grid operational regime in China's power sector gives rise to strong institutional inertia. The flexible economic dispatch of coal power plants, which could support higher renewable consumption, is economically and operationally not viable under the current regulatory regime. The prices in medium- to long-term electricity contracts, which make up the bulk of all traded volumes in China are heavily regulated with price ceilings and floors set by the NDRC (at ±20% in 2024), limiting flexibility. Spot market prices are similarly controlled. Retail prices are fixed by provinces year ahead, based on the benchmarks set by the NDRC, and are subject to the final approval by the NDRC[14,15]. However, coal prices are floating, rendering coal power plants to operate at a loss at times when the cost of operation is higher than the fixed electricity price, causing supply issues[16,17]. In addition, the fixed prices in medium- to long-term contracts in turn pre-determines generator revenue and scheduling, usually a year ahead of actual delivery, aligned with the current regulatory regimes under which the dispatch operators prioritize grid stability and outage prevention over economic efficiency and flexibility.

On the other hand, there are also factors that drive long-term shifts towards renewables. For example, coal generation has become more expensive in recent years due to higher energy prices in general. Renewables have become extremely cheap in contrast. Typical levelized cost of electricity (LCOE) for PV in the coastal provinces of China is around 0.03$/kWh, whereas for coal thermal power plants it is around twice as much (0.05-0.06$/kWh)[18,19]. Not only is solar PV the cheapest source of energy everywhere in China[20], the LCOE for solar PV combined with lithium batteries has also reached parity with the running cost of coal power plants during peak demand hours in certain regions[21]. Compelled by economic factors, a growing share of energy intensive industries such as photovoltaic panels, carbon fiber, polypropylene, synthetic ammonia, and primary aluminum are already moving from the east to the west due to the much cheaper energy costs there[22–24].

In the near term, the regulators of the power sector have taken steps to respond to these new developments, but the trajectory of its transition remains unclear. Recent reports highlight significant challenges in reforming market rules and pricing systems, including steep learning curves and political economy constraints at national and local levels[25,26]. These issues manifest in grid stability concerns during extreme weather, fears of supply shortages, power plant cash flow problems, difficulties balancing supply with renewable intermittency, and lack of clarity over the new price structures. Although pilots for spot markets and interprovincial trading have been introduced, their small scale and exclusion of long-term coal contracts haven't significantly improved system flexibility. Slow reforms risk continuing the market regime where the spot market is limited in size and is under tight price controls, leading to negative prices and discouraging renewable investment[27]. On the other hand, fast reforms aimed at full efficiency without regulatory changes regarding the grid could strain market liquidity and even cause power outages, risking popular acceptance to reform[17]. With a high renewable build-out target but slow reform pace, renewable curtailment rates are rising[28,29]. Given the current low renewable share in China (around 15% compared to 55% in Germany), these curtailments highlight inefficiencies in the current system. Continued coal plant build-out, lack of market flexibility, overcapacity, and rigid grid management indicate institutional inertia, likely causing multiple missed targets in the 14th five-year plan (FYP)[30].

Given both uncertain coal phase-out and the relatively more certain expansion of direct electrification, a major question remains, namely, *will the faster pace of electrification increase China's emission when the power sector could continue to be dominated by coal? What are the overall and sectoral climate impacts of a slower demand-side direct electrification vs. a slower coal phase-out?* Existing literature based on "bottom-up" emission analysis has focused on specific technologies such as EV and heat pumps[31–33], however, in the case of China, there is no systematic study from a whole energy system perspective.

We bridge the gap in existing literature through several innovative aspects. First, by using the global integrated assessment model (IAM) REMIND, we contextualize the study of a single-country high-emitting sector within global modeling. This links China's sectoral transition to global peak temperature projections and the broader mitigation landscape (see Methods). We chose REMIND for its proven success in IPCC vetting processes, contributing to the highest number of scenarios in the latest IPCC assessment (AR6)[34]. The model is also a core contributor to the Network for Greening the Financial System (NGFS) project with central banks worldwide[35,36]. REMIND's stylized power sector model aligns with recent studies using high temporal and spatial resolution (e.g., hourly data and provincial-level resolution) to simulate high shares of solar and wind energy[37–39], while also reproducing current prices (see Figure S3). These features make REMIND uniquely suited to modeling China's climate mitigation.

Second, using the latest developments of a state-of-the-art IAM, we can accurately analyze the aggregate emissions and carbon footprints of emerging and incumbent end-use energy technologies. This enables us to map detailed sectoral pathways to cumulative emissions and their impact on peak temperatures, facilitating a thorough discussion of the interaction between power sector transformation, electrification, and sectoral emissions from a systems perspective. Beyond emissions, the high electrification REMIND-China scenarios align with the latest technology trends, providing self-consistent projections for primary and secondary energy mixes, sectoral energy use, power price development, and end-use applications like EV scale-up and steelmaking. Notably, the high electrification, coal phase-out scenario envisions an energy system where nearly all solid fuels are phased out by net-zero, except in the cement sector, representing a radically different future for energy generation, transport, and consumption in China.

Third, IAM REMIND allows for detailed modeling of technologies and industrial policies in demand-side sectors. For example, policies supporting electrification, such as EV charging infrastructure and heat-pump subsidies, can be integrated into the broader policy package. This approach is especially relevant given the success of targeted green industrial policies in China. The scenario assumptions take advantage of this feature to incorporate high shares of direct electrification, particularly in the transport and building sectors (see Table 2 and Methods). Due to these granular features, this high-electrification scenario is unique among other IAM scenarios with similar climate constraints in the IPCC AR6 database (see Figure 3)[40]. Our study also includes GHG emission analysis of China in a global context (Supplemental Material 5) and scenarios with a fixed budget for China's contribution to peak temperature (Supplemental Material 9), which are not commonly found in many China IAM scenarios, where $CO_2$-only analysis and net-zero neutrality years are often used[29].

Based on detailed demand-side pathways and sectoral emission-impact analysis, we derive policy implications. We show that rapid electrification and efficiency improvements are essential, robust, and effective for emission mitigation, leading to significant medium- to long-term $CO_2$ reductions. We also emphasize that through efficiency improvement, electrification of industry does not necessarily create vastly more electricity consumption. Our findings align with recent studies, such as Yin et al. (2023), which use an extended Kaya factor method to assess emission reductions from electrification and electricity decarbonization in the U.S., EU, and China[42].

# Results

The scenarios in this study are divided into two groups. The first group includes four reference scenarios with no additional emission constraints, except for moderate existing regional carbon prices. In these scenarios, the economy-wide electrification rate—defined as the share of electricity in the final energy mix—only increases to 40% from today's 30%, then stagnates ("El-lo" for "low electrification"), reflecting a baseline for moderate electrification. There are no additional industrial policies (i.e., non-$CO_2$ price policies) supporting electrification. The second group consists of four 2060 net-zero mitigation scenarios for China, modeled under a global carbon budget consistent with the well-below 2°C target and regional net-zero goals for the major economies, namely EU, the US, China and India. These scenarios are summarized in Table 1.

In addition to the $CO_2$ price policy, which arises endogenously from the emission constraints, we include additional policy levers to promote direct electrification, such as subsidies for heat pumps and support for electric vehicles. This results in high electrification dynamics ("El-hi"). The policy scenarios, combining $CO_2$ pricing and electrification-promoting policies with baseline electrification dynamics, are summarized in Table 2. For each sector, the impact of prices (e.g., energy and $CO_2$ prices) may differ due to its unique transformational dynamics. For instance, in the transport sector, industrial policies have a stronger influence, as the adoption of EVs depends more on infrastructure development and consumer acceptance than solely on fuel prices.

| Scenario group | Climate constraint | Coal power pathway | | | |
|---|---|---|---|---|---|
| | | 2025 peaking at a high level, phasing down in 2030, 2060 phase-out | 2025 peaking at a medium level, 2030 phasing down, 2050 phase-out | 2025 phasing down, 2045 phase-out | 2025 phasing down, 2040 phase-out |
| Current policy (results in low electrification / “El-lo”) | No emission constraint. Existing $CO_2$ prices are assumed for future years | **El-lo CP-NPi** | **El-lo CP-plateau30** | **El-lo CP-medium** | **El-lo CP-fast** |
| Climate policy (results in high electrification/ “El-hi”) | Global cumulative $CO_2$ emission (2020-2100) less than 1150 $GtCO_2$ (> *67% likelihood for* below 2°C warming). China $CO_2$ emission reaches net-zero in 2060, U.S. and EU 2050, and India 2070. Endogenous $CO_2$ price determined by climate constraint | **El-hi CP-NPi** | **El-hi CP-plateau30** | **El-hi CP-medium** | **El-hi CP-fast** |
| Climate policy (results in high electrification/ “El-hi”) with regional emission budget constraints “reg budget”, **see Supplemental Material 9** | Same as “Climate policy” above, in addition: impose a 210 $GtCO_2$ emission budget on China between 2020 and 2060 | **-** | **El-hi CP-plateau30 reg budget** | **El-hi CP-medium reg budget** | **El-hi CP-fast reg budget** |

*Table 1: The two-dimensional scenario design includes four reference scenarios (“low electrification / El-lo”) and four mitigation scenarios (“high electrification / El-hi”), with scenario names in bold. It outlines emission constraints, corresponding $CO_2$ price trajectories, and the four levels of coal power phase-out pathways. The relationship between each scenario’s assumptions and its eventual electrification rate is detailed in Table 2. The last row of scenarios, where China’s regional emission constraint is fixed, is only shown in Supplemental Material 9.*

Within each scenario group, we vary pathways for unabated coal power capacity and capacity factors, creating four sub-scenarios for the coal power phase-out timeline (Table S1). Table 1 summarizes these pathways, which are detailed in Table S2 and visualized in Figure 1B. For the medium- to long-term, the coal phase-out pathways reflect the uncertainty range in leading domestic models[41,43] and are comparable to historical phase-out speeds, including the rapid pathways in IPCC scenarios, which have faced recent feasibility concerns[44–46]. These four paths represent a 5-20 year uncertainty range, all aligned with the net-zero pledge. We derive three main results from these scenarios, outlined in the following subsections.

On the other hand, the sectoral electrification outcomes of the scenarios depend on a more complex set of assumptions in the model, and depends on four main aspects (see Table 2): 1) Baseline dynamics consists of ongoing electrification due to economic development, such as increasing household appliances and cooling use; 2) Endogenous electrification via climate constraint, i.e., the $CO_2$ prices induced from the climate constraint influence both the speed of near- and long-term electrification rate; 3) Industrial policies which are exogenously imposed policies, such as subsidies or taxes, further encourage electrification; 4) Impact from various speeds of coal power phase-out, which imposes a mild influence on electrification, mainly through electricity pricing changes resulting from different speeds of coal power phase-out.

| | | Current policy / reference scenario ("El-lo") | China net-zero scenario ("El-hi") | |
|---|---|---|---|---|
| **Sector** | **Sub-module modeling methodology** | **Impact of main baseline dynamics and assumptions on sectoral electrification** | **Endogenous sensitivity of sectoral electrification towards price mechanism** | **Additional exogenous assumptions for industrial policies supporting electrification** |
| **Transport** | Logit-based formulation, iteratively soft-coupled to optimization based energy system | Relatively weak future support around charging infrastructure, more hybrid vehicles due to range anxiety. Policy scenario "Mix3" is used due to the recent optimistic technology development in China (see Methods) | Relatively low sensitivity towards endogenous $CO_2$ price or electricity price | Support policies around charging infrastructure, innovation on batteries shortens charging time. Policy scenario "Mix4" is used for high ambition scenario (see Methods) |
| **Industry** | Industrial process-based formulation for steel; Final energy mix determined via "CES"-based formulation for cement, chemicals and other industries | Baseline dynamics result in a moderate increase in the electrification rate due to: 1) the assumed slow decline in cement and steel output as infrastructure and building stock saturate; 2) the baseline increase in energy efficiency of industrial products, in line with historical trends; and 3) the cost advantage of secondary steel over primary, leading to a shift from primary to secondary steel production, which boosts the electrification rate (the share of primary to secondary steel production is capped to reflect that not all steel demand can be met by scrap steel[47]) (see Figure S10). | High $CO_2$ price and low electricity price drive energy efficiency and a shift to direct electrification (see Supplemental Material 8). This combination of lower overall sectoral energy demand and slightly higher electricity demand in industry leads to a higher electrification outcome | Subsidy for high temperature heat in other industries (250$/MWh(el)), tax on hydrogen application in cement (150$/MWh(th)), other industry(70$/MWh(th)) and chemicals (20$/MWh(th)) to encourage direct electrification |
| **Buildings** | Service demand is generated via simulation-based model; Final energy mix determined via "CES"-based formulation | The electrification rate naturally increases as assumed in the baseline: 1) The demand for useful energy from electric appliances, lighting, and space cooling grows with GDP per capita (no saturation); 2) the adoption of air conditioning also rises with GDP per capita but saturates at high levels of GDP per capita. | High $CO_2$ price and low electricity price increase the electrification rate in buildings through heat pumps, replacing solid biomass, coal, and gas as heating sources. | Heat pump subsidy (80$/MWh(el)) for the household, tax for district heating (25$/MWh(th)) |

*Table 2: The relation between the scenarios' policy assumptions and the sectors' scenario outcome in terms of electrification rates. We make clear the three main underlying dynamics that influence direct electrification in the scenarios: 1), baseline dynamics where only current policies are assumed (column 3); 2), the impact of $CO_2$ price due to climate constraint as well as the impact of electricity price due to coal phase-out on shifting the sector towards direct and indirect electrification (column 4); 3), additional subsidies used for further supporting direct electrification technologies (column 5).*

## Climate impact of sectors under various electrification and coal pathway scenarios

Based on scenarios with varied coal pathways and electrification in the two-dimensional scenario space, we analyze the interdependencies of these factors and derive climate impacts from the sectors. As Figure 1 shows, in the mitigation scenarios, long-term electrification rates are not strongly affected by power sector emission intensities in the REMIND-China mitigation scenarios. A high direct electrification rate under the climate constraint is a robust feature of China's decarbonization, reaching around 60% before 2055 with additional industrial policies (Figure 1A).

However, there are significant climate effects from the power sector itself, as well as uncertainties in its transition. Figure 1D, 2 and Table S4 detail the impact of China's power sector on peak temperature warming. Even in the highly optimistic scenario ("CP-fast"), the cumulative emissions from the Chinese power sector alone are around 39 $GtCO_2$ between 2023 and 2060. This is about 26% of the remaining 1.5°C global carbon budget of 150 $GtCO_2$ (67% likelihood of keeping warming below 1.5°C, or 250 $GtCO_2$ for a 50% likelihood), and 4.1% of the remaining well-below 2°C target of 950 $GtCO_2$ (67% likelihood). In the slower "CP-NPi" path, the power sector's cumulative emissions reach 99 $GtCO_2$, or 66% (10.4%) of the global 1.5°C (2°C) budget, with more than half emitted after 2030. As shown in Figure 1C and Table S4, compared to the optimistic "El-hi CP-fast" scenario (143 $GtCO_2$ between 2023 and 2060), a ten- or fifteen-year delay ("El-hi CP-plateau30" or "El-hi CP-NPi") in coal phase-out adds around 24% (35 $GtCO_2$) or 39% (56 $GtCO_2$) more to economy-wide cumulative $CO_2$ emissions from 2023 onward. The power sector impact is equivalent to 0.02°C additional warming for roughly every 10 years of delay in coal power phase-out (0.18°C difference between temperature impact from power sector for "El-hi CP-fast" and "El-hi CP-plateau30", or 0.19°C difference between "El-hi CP-medium" and "El-hi CP-NPi").

Compared to the current policy path with a low electrification rate ("El-lo CP-NPi"), even the slow coal phase-out path under high electrification ("El-hi CP-NPi") can abate around 116 $GtCO_2$ of cumulative emissions. Of this, 70 $GtCO_2$ can be attributed to more energy-efficient and electrified end-use sectors, while the remainder is linked to non-energy sectors, such as land use, process emissions, extraction, and waste. This underscores the importance of electrification and energy efficiency in the demand sectors for a successful energy transition in China, as this level of cumulative abatement is equivalent to a roughly 15-year faster power sector transformation.

Using "Transient Climate Response to cumulative Emissions of carbon dioxide" (TCRE), we can translate between the cumulative emission and temperature degree warming. This factor can be taken as roughly 0.45°C per 1000 $GtCO_2$ of cumulative emission (uncertainty range 0.27–0.63°C, as given by the latest literature[48]). Using the TCRE we can estimate the role specific sectoral pathways play in lowering world peak temperature (Figure 2).

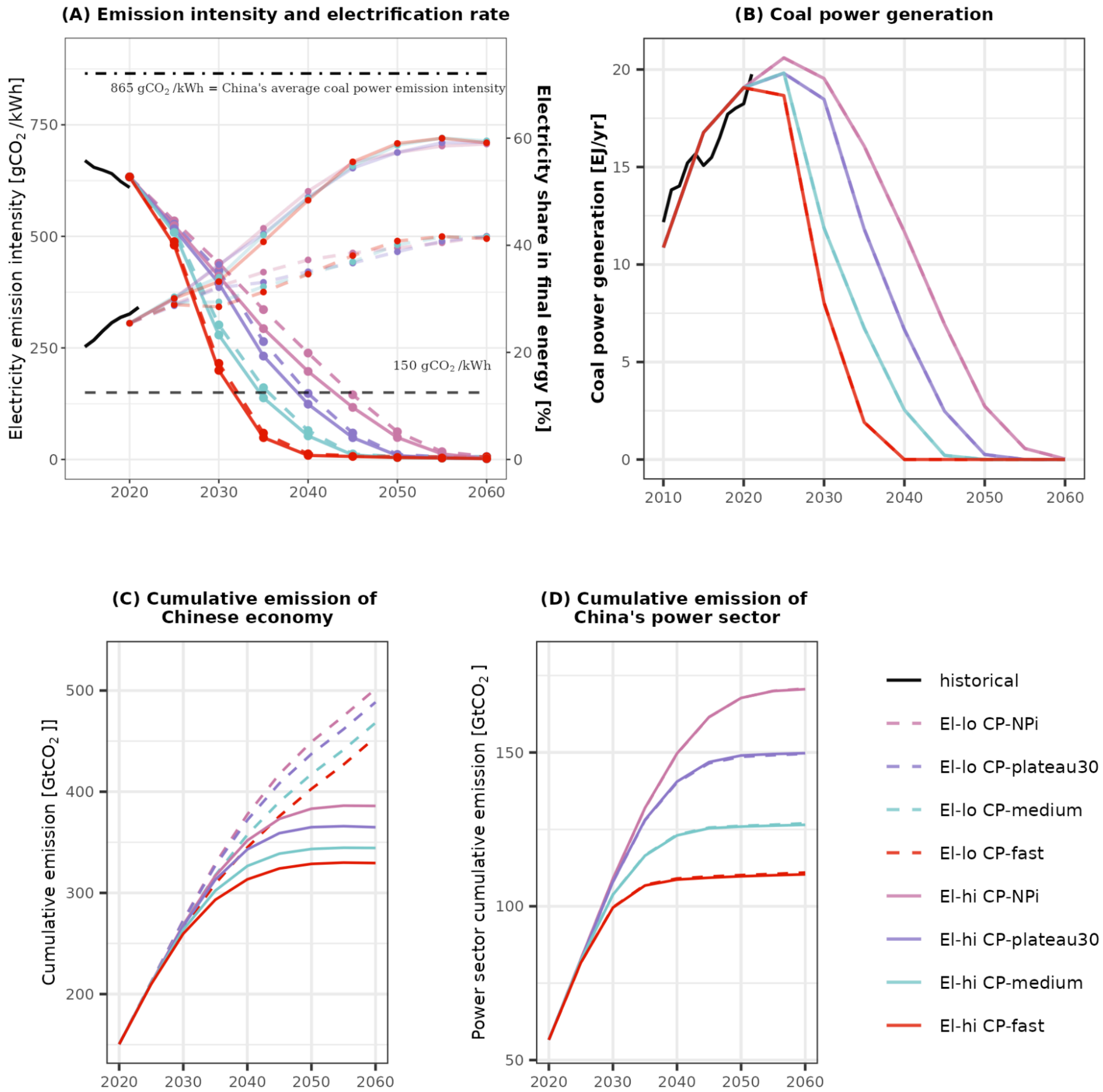

*Figure 1: China's climate impact under coal power phase-out uncertainty and levels of electrification is illustrated as follows: (A) Electricity emission intensity (left axis) and end-use electrification (right axis); (B) Annual coal power generation; (C) Economy-wide cumulative emissions (since 2005); (D) Power sector cumulative emissions (since 2005). In (A), the average coal power emission intensity in China is shown, along with a low electricity emission intensity threshold of 150 g$CO_2$/kWh for reference. The former uses the "13th FYP $CO_2$ equivalence of binding target for average coal consumption of operating coal-fired units", collected by the IEA statistics[49]. The historical power emission intensity data is obtained from Ember, noting that due to REMIND resolution, an aggregated R10 CHINA+ region is used. Historical data for electrification rate is sourced from the IEA. For a sub-national distribution of emission intensity, see Figure S17.*

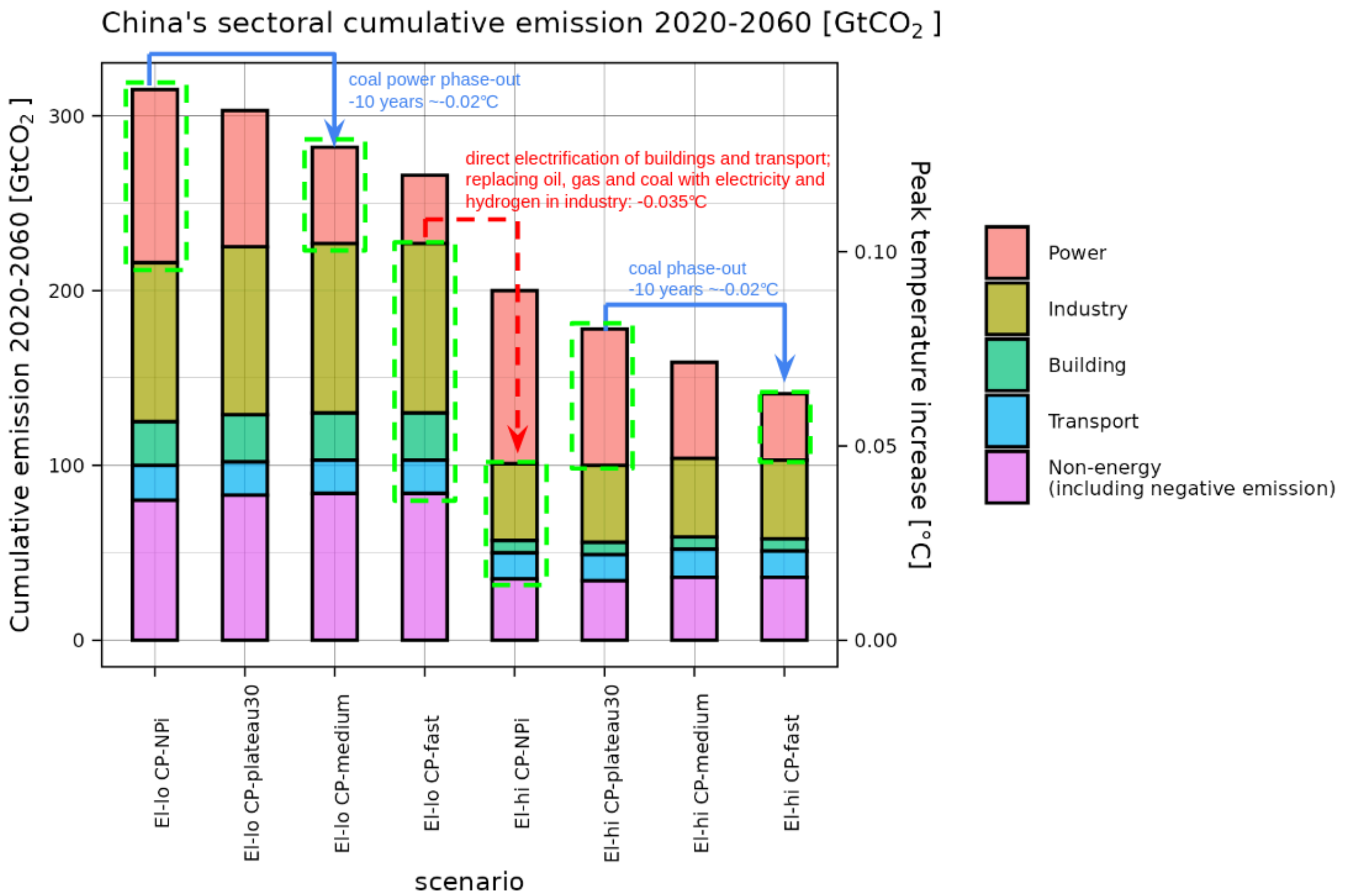

Figure 2: Cumulative emission of China's sectors until net-zero year and their corresponding impacts on peak temperature increase.

Figure 2 shows the cumulative emissions from China's energy sector and their impact on peak warming. It is clear that cumulative emissions are reduced through two almost independent factors: the supply and demand sides of the energy system. The supply-side dynamics are driven by the timing and path of coal phase-out, while demand-side transformation is led by the replacement of the direct use of coal, gas and oil in the demand sectors. The latter means a combination of direct electrification of steel, moderate electrification of cement and chemicals industries, increased hydrogen use and energy efficiency improvements in heavy industries. Either option – 15 years of faster coal phase-out or demand-side efficiency and electrification, corresponds to about 0.035°C of avoided global warming.

Figure 2 also clearly shows that a slower coal phase-out combined with fast electrification of demand sectors does not result in higher cumulative emissions in the power sector. Instead, it highlights the significant benefits of an electrified end-use sector. If China phases out coal power by 2055 without addressing end-use sectors ("El-lo CP-NP"), it will contribute to approximately 0.142°C warming (315 GtCO2 cumulative emission). However, if China phases out coal power by 2055 and also increases energy efficiencies and electrification ("El-hi CP-NP"), it will contribute to roughly 0.09°C warming (199 $GtCO_2$ cumulative emissions), and if combined with a rapid coal phase-out ("El-hi CP-fast"), it will contribute to about 0.064°C warming (143 $GtCO_2$ cumulative emission). Consequently, as long as China reduces power sector emission intensity to at least 150 $gCO_2$/kWh in the next 15 years (all scenarios except "CP-NPi" reach this by 2040), rapid electrification offers significant emission abatement potential in the medium- to long-term, equivalent to avoiding at least 0.044°C of global warming. This result comes from combining 0.01°C abatement from the power sector between the "El-hi CP-NPi" and "El-hi CP-plateau30" scenarios, and 0.034°C demand-side (sum of non-power sector) abatement between the “El-hi” and “El-lo” scenarios for the “CP-plateau30” path.

In addition to scenarios under the net-zero 2060 constraint for China, we also compute scenarios under fixed national emission budget constraints (Supplemental Material 9). These results show that if China’s contribution to warming is fixed to its roughly equal per capita remaining budget allocation – 210 $GtCO_2$ between 2020 and 2060 – a ten-year delay in coal phase-out would nearly double negative $CO_2$ emissions from around 0.6 $GtCO_2$/yr to 1.2 $GtCO_2$/yr by 2060. Using emission budget accounting from 2023 onward, we highlight the immense uncertainty in China’s power sector emission pathway. However, it is important to note that this method of measuring regional contributions to the global budget does not consider historical emissions or global structural inequality (Supplemental Material 10).

Other basic results related to secondary electricity mix and primary energy mix are in Supplemental Materials 2-3 (Figure S1 and S2). Selected results in buildings, transport and industry sectors, such as final energy (FE) sales and stock of passenger EVs as well as steel sector transformation, are listed in Supplemental Material 7 and 8 (Figure S7-12).

## Sectoral electrification rates differ in response to price signal vs. regulatory policies

Economy-wide end-use direct electrification under climate constraints is robust at a system level, regardless of coal power phase-out speed. However, the response to power sector dynamics slightly differs. Electrification rates respond to prices, but this effect does not dominate in the long run. There is, however, some path dependency. As Figure 3 shows, the speed of direct electrification moderately correlates with the pace of power sector mitigation in the near to medium term. This is expected, as sectoral electrification rates in REMIND react to electricity prices, which peak higher during faster mitigation actions (see Figure S3). Therefore, there is a slight near-term trade-off between supply and demand: delays in coal power phase-out encourage near-term electrification slightly (by 3.2% in 2030 between “El-hi CP-plateau30” and “El-hi CP-fast”) because power prices are lower due to existing coal plants' lower production costs. However, long-term electrification is slightly discouraged (by 1.7% in 2050 between “El-hi CP-fast” and “El-hi CP-plateau30”) due to higher long-term power prices from slower learning effects in renewable energy generation.

Due to the ways different sectors are modeled in REMIND, different sectors’ electrification rates react differently to the power sector economics. Industry and building sector rates react to the power sector transition speeds slightly stronger than the transport sector, reflecting lower price-sensitivity of the latter (see Table 2 and a detailed description of sector modelling methodology in the Methods section). Energy costs matter much more for the production costs of basic materials in industry and heat provision for the building sectors, giving rise to relatively greater than transport, though still small price response.

In general, the sectoral direct electrification rates of REMIND China scenarios are higher than other IAM scenarios in the AR6[50]. Except industry electrification rate, for which REMIND is slightly lower than the average AR6 scenario, buildings and transport electrification rates in the REMIND scenarios are much higher, due to the assumptions of high electrification in the demand sectors in line with technology trends (see Methods). This results in around 5-10% higher rate of the whole economy than most AR6 well-below 2°C scenarios with equivalent climate constraint (Figure 3 top-left panel).

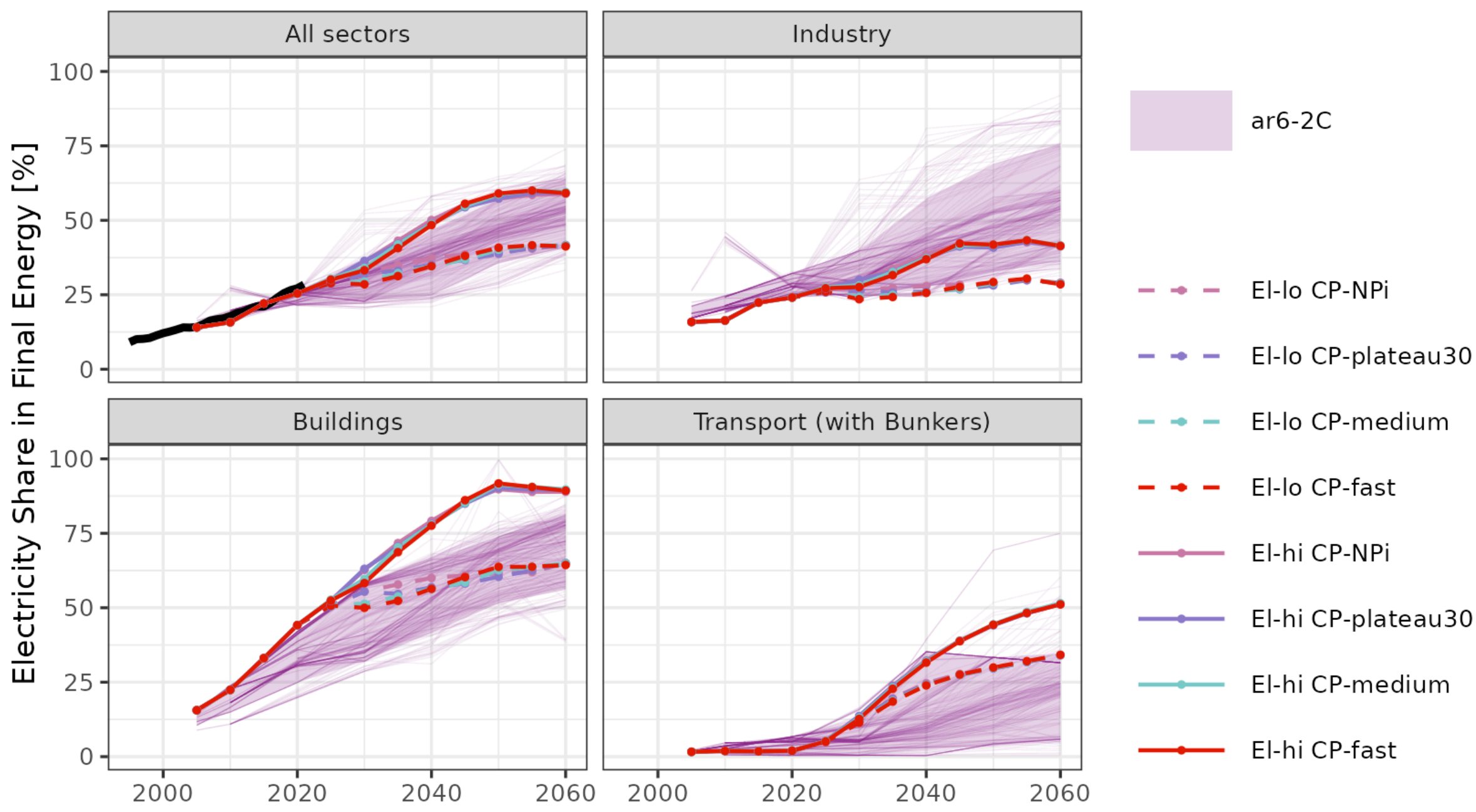

*Figure 3: Economy-wide electrification rate and the electrification rates of the three end-use sectors under the 8 scenarios, compared to IPCC AR6 scenario data. The thin lines indicate all scenarios, and the darker area indicates the 10-90% quantiles of these scenarios for 2°C[50]. For the R10 CHINA+ regions, the regional scenario C3 is selected for the well-below 2°C scenario (> 67% likelihood for below 2°C warming).*

## Comparing emission intensity of electrification vs. incumbent end-use energy technologies

To demonstrate the reason that rapid scaling up of electrification technologies does not lead to increased emissions, we assess the emission impacts on multiple scales for sector-specific applications, as shown in Figures 4 and 5. In Figure 4, we compare the emission intensity of incumbent fossil-based technologies and direct or indirect electrification technologies. We find that emission intensity parity is reached within different time frames depending on the energy conversion efficiencies of the applications. For three significant use cases—electric arc furnaces (EAF) for secondary scrap steel production (Figure 4A), heat pumps (Figure 4B), and battery electric vehicles (BEVs) (Figure 4C)—their superior efficiencies already result in equal or lower emissions than current fossil-based applications, even with China's current electricity mix. For other applications, such as resistance heating steam boilers or electrolytic hydrogen-based steel making (H2-DRI-EAF) where the electrolysis is not flexible (i.e. the power intensity is the average grid intensity at all times), emission intensity parity is expected in the late 2030s, but can be advanced by 4–6 years under faster power sector transitions (Figure 4A). However, these electrification applications remain relatively niche in China (compared to coal-based heating CHP and coal-based primary steel-making BF-BOF), meaning that, on an aggregate level, we do not observe emissions increases from rapid electrification. The data for heat applications are sourced from Agora Energiewende[51] and IPCC[52].

In Figure 4, we note that "power emission intensity" on the x axis, as is elsewhere in this paper, reflects that of the final energy electricity after grid losses, and not the LCA carbon footprint of electricity. The LCA carbon footprint of electricity generation, while part of the input to the LCA algorithm, is not shown here for simplicity and comparability across service types (see LCA Module in Methods section). The LCA carbon footprint of electricity is usually 20% higher than the direct emission intensity of power generation, due to the additional embedded "indirect" emissions in power production, i.e. those upstream emissions resulting from the production of solar panels or the construction of wind turbines. The detailed prospective LCA calculation for the carbon footprint of electricity is not shown here due to limited scope, but the method is described in workflow diagram S21.

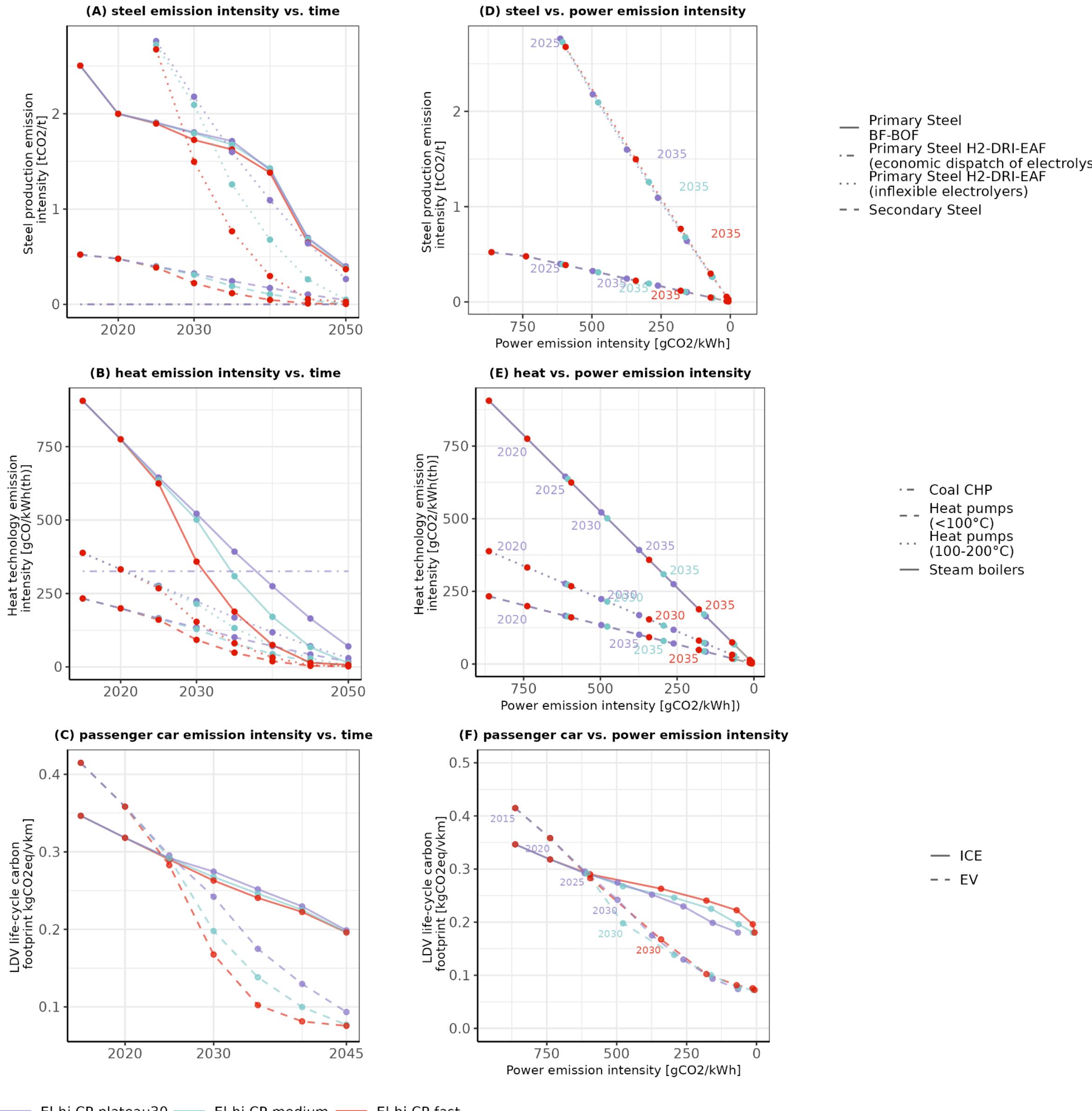

*Figure 4: The comparison between emission impact of the electrification of three energy service applications—steel production, building heating, and light-duty passenger (LDV) vehicles in transport —with that of the incumbent fossil options. Left column: comparison in terms of emission intensity (or carbon footprint for LDV) per energy service unit over time. Right column: emission intensity (or carbon footprint for LDV) per energy service vs. power emission intensity. Emission intensity parity is identified by the intersection of the electrification and incumbent trajectories. The example of indirect electrification using electrolysis–DRI H2 steelmaking (A) is not actually assumed in the model, and is only shown here to highlight the role of economic dispatch of electrolyzers, which,*

*when charged with low-cost intermittent renewable power, is assumed to have zero emission intensity in the model.*

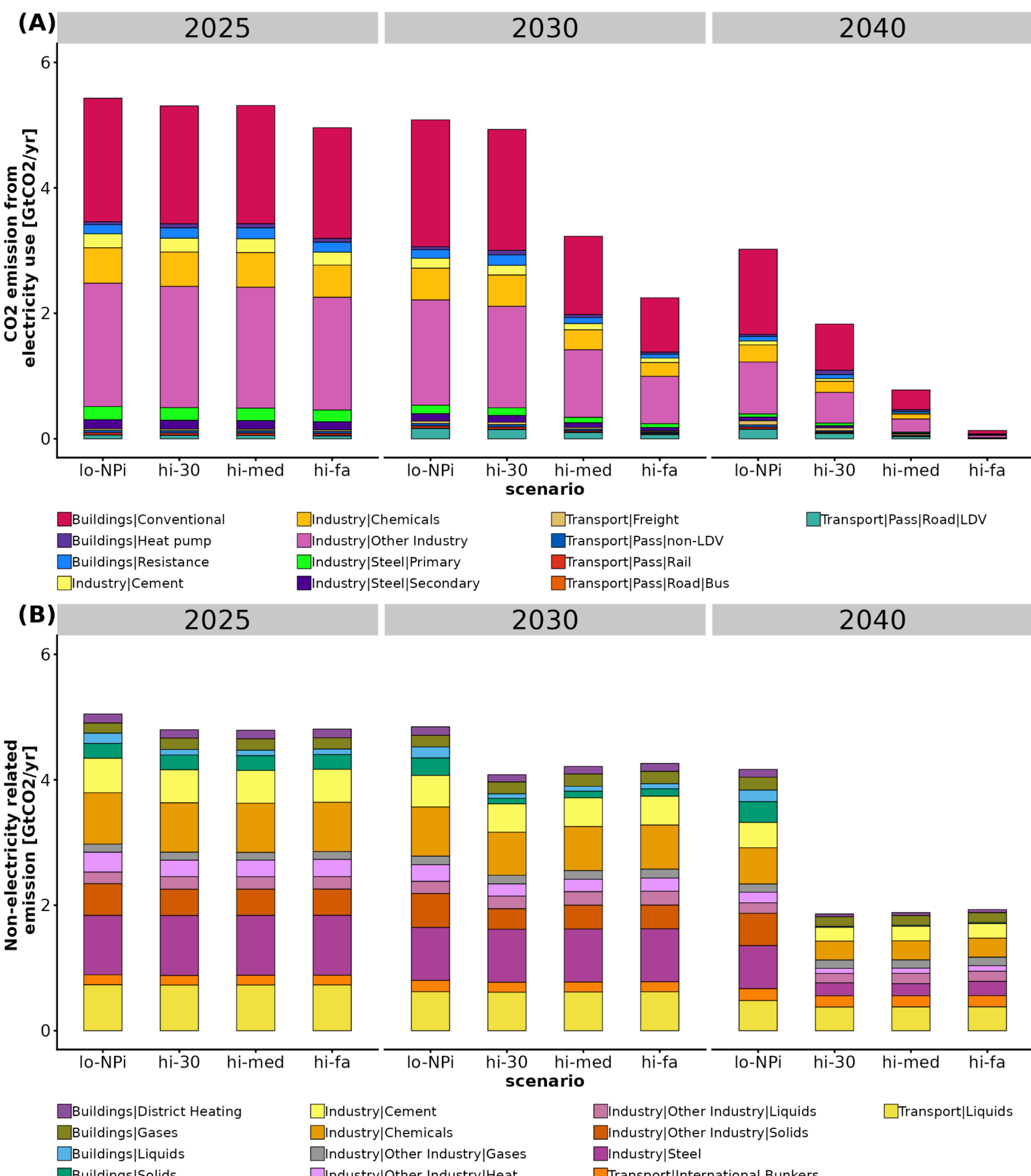

*Figure 5: Annual* $CO_2$ *emissions from electrified vs. non-electrified buildings, industry, and transport subsectors. For clarity, we focus on a subset of scenarios: "El-lo CP-NPi" ("lo-NPi"), "El-hi CP-plateau30" ("hi-30"), "El-hi CP-medium" ("hi-med"), and "El-hi CP-fast" ("hi-fa"). Emissions from secondary energy sources (solids, liquids, and gases) due to conversion losses are not shown here. Note that 2025 is an average year between 2023 and 2027. The emission shown here does not include non-*$CO_2$ *emissions.*

Compared to current internal combustion engine (ICE) vehicles, BEVs' carbon footprint only reaches parity in 2024, due to the energy-intensive production of lithium batteries[53]. Significant geographic variations in emissions exist within China, particularly in North and Northeast China, where power emission intensity remains around 800 $gCO_2eq/kWh$[54]. This results in notable indirect emissions from early electrification in these regions, especially if EV batteries and cars are produced and used locally[55]. A province-level LCA would require supply-chain location data, particularly for battery manufacturing.

In cases where early electrification may lead to increased emissions, there is no one-size-fits-all answer on whether it's always detrimental for the climate. While the period of higher indirect emissions, caused by slower coal phase-out, may last several years, it is short compared to the long timescales required for technological diffusion, stock replacement, and learning. Many direct electrification technologies currently have small market shares, but under optimal pathways, electrification rates could reach nearly 50% by 2040, up from 27% today. For instance, while passenger EVs could achieve 90% sales share by 2100, diesel cars would still be in use until 2040. Therefore, gradual stock replacement means electrification should begin now, rather than waiting for a fully clean grid, to avoid further fossil fuel lock-in.

On the aggregate level for each subsector, there is no evidence for sudden short-term emission increases from switching to electrification technologies. Strikingly, as Figure 5 shows, even though the electrification rate across all sectors is currently low, within the total annual $CO_2$ emission, around 5.33 $GtCO_2$ comes from already electrified services, which consist of non-heating demand in buildings, other industry such as aluminum (0.67 $GtCO_2$[56]) and paper-making (~0.14 $GtCO_2$[57]), secondary steel, as well as mechanical and low-temperature heat in chemical sectors. Non-heating conventional electricity usage in the building sector consists of space cooling, lighting, electric appliances and miscellaneous equipment. The energy services provided in these already electrified sectors have emission intensities that are essentially proportional to the power emission intensity. This is why comparing across the scenarios, we see the biggest emission saving potential from rapid greening of the power supply for the industry and buildings sectors, usually at a rate of 1-2 $GtCO_2$ per decade.

The emission saving in electricity related sectors includes a reduction of emissions from already electrified sectors as well as avoided emission of newly electrified services, in addition to a small amount of electrification induced indirect emission increase. Nevertheless, the last category plays only a minor role due to the aforementioned slow stock-replacement, and is not significantly visible in Figure 4. For example, even under the scenario of no coal phase out until 2030 (“plateau30”), in 2030, compared to 2025, the total road BEV results in an additional annual indirect emission of 94 $MtCO_2$ for passenger BEV, 1.6 $MtCO_2$ for electric buses, 8 $MtCO_2$ for passenger non-LDV, and 12.5 $MtCO_2$ for electrified freight in the year 2030. The additional annual emission from electrifying road transport totals to around 116 $MtCO_2$ in 2030, which is slightly larger than the abated direct emission from oil use in transport in the same year (around 114 $MtCO_2$ comparing 2025 to 2030). However the technology advances that happen in the decade between 2020-2030 leads to large abatement down the road when the stock replacement is completed (i.e. annual 500 $MtCO_2$ abatement from reducing transport oil demand in 2050 compared to 2025). So while we can quantify a temporary surplus annual emission of 2 $MtCO_2$ in 2030 from rapid electrification in transport combined with a slower coal phase-out schedule, this is a small portion of total national emission to be abated (around 10 $GtCO_2$ per year). As long as competitive electrification technology results in complete or almost complete stock replacement later on, it is a worthwhile endeavor despite that the power mix is not entirely low-carbon.

We note that in REMIND v3.3.0, we assume electrolytic hydrogen only runs during hours where renewable is abundant, and therefore has almost zero emission as long as its dispatch is assumed to be economic (see Methods under "Power sector module"). This assumption is in line with the current development in China, with new hydrogen projects located in renewable abundant western regions, and using completely renewable electricity for production[58–62].

One should note that due to the slow stock-replacement rate of non-electric technologies with electrified ones, the largest visible abatement of direct emission only happens between 2030 and 2040 for all high electrification scenarios. While the "CP-fast" scenario gives a steady 2 $GtCO_2$ per decade abatement pace in the power sector, only until 2040 does this translate to 2 $GtCO_2$ per decade of abatement in the non-electrified sectors. This is due to the multiplicative effect of the strategy of decarbonizing the power sector and simultaneous pushing for direct electrification. This highlights the importance of early adoption of technologies which have slow stock replacements. The investment today in emission abatement technologies can take two decades before an eventual large reduction of emission.

## Discussion

To achieve the lowest possible temperature rise, global and national emission pathways should be as convex as possible, emphasizing rapid near-term emission reductions. This is particularly crucial for China, given its high current emissions. Our sensitivity analysis suggests that both coal power phase-out and electrification must be supported by policies under stringent climate goals.

Our analysis shows that, for China, under all scenarios examined, there is no significant extra transient emissions from early diffusion of electrification. This is mainly due to two factors. First, among the three demand sectors, only road transport and buildings have increasing electricity demand (Figure S1), and when they do, they replace the direct use of coal, oil or gas. Industry's electricity demand remains relatively constant, despite its share rising by 12% in final energy, due to a slight decline in output and improvements in energy efficiency. Thus, the higher electricity share in final energy does not necessarily indicate increased electricity demand in industry. The main electrified sub-sector in industry is steel, where electric arc furnaces are much more energy-efficient and have a lower carbon footprint than primary steel. The little increase in electricity demand combined with the high efficiency of electrification technology leads to low chance of transient emissions from electrification.

Second, while electrification and absolute electricity consumption are rising in buildings and transport, these sectors have slower adoption rates due to their unique characteristics. For example, electric vehicles require time to diffuse through the market, with increasing sales, infrastructure development, and heterogenous decision-makers. As long as electricity emission intensity becomes low in the long term, early and rapid adoption of technologies in these sectors will not cause significant additional emissions. These two sectors are also by far less energy consuming than the industry sector, resulting in less absolute indirect emission increase due to their lower shares of China's energy demand overall.

Using in-depth multi-level analysis of aggregate and specific emission drivers, this paper challenges the strictly sequential interpretation of the "order of abatement", which suggests that direct and indirect electrification should only follow after the power sector transformation is nearly complete. In fact, even with coal generating half of China's power, if the average power emission intensity drops below 150 $gCO_2/kWh$ in the next 15 years, a rapid scale-up of electrification in steel, road transport, and buildings, combined with efficiency gains, can lead to significant emissions reductions. This could be equivalent to a 15-year earlier coal phase-out (both are around 0.035°C). A more nuanced understanding of the “order of abatement” considers not only the marginal abatement cost curve, but also factors like end-use emission intensity based on electricity emission intensity, turnover limits of non-electrified products, and aggregated indirect emissions from newly electrified sectors.

We summarize three main policy conclusions and future research directions based on our analysis: First, in China, policy research should focus on decarbonizing the power sector through understanding the complex price structure of a system with significant renewable shares. Regulations should prevent dividing the spot market into “regulated” and “liberalized” parts, which can distort prices and cause negative prices even when renewable share is low. To avoid power outages during electrification, capacity-based compensation for coal plants may be needed to prevent premature closures, especially as they remain essential for peaking services. However, capacity payments should be tied to flexibility requirements. Technological advances in big data and AI can enhance grid efficiency, helping China’s grid operators adopt state-of-the-art operations. This is especially advantageous, given that AI algorithms used for grid operation today do not involve large power consumption and are distinct from large language models which are power intensive[63,64]. Regardless of the reform approach, early institutional learning is crucial. The uncertainty around China's electricity sector reform calls for collaboration between energy modelers, economists, and grid experts to design effective policies.

Second, early adoption of end-use electrification should receive near-term policy support, guided by application efficiency and local grid emission intensity. Historically, environmental policies have included campaign-style measures, such as regulatory bans on sales and stocks[65–67], leading to quick stock replacements in specific regions, like Beijing's coal boiler replacement before the Olympics or truck bans in cities. There is a small risk that electrification may occur “too soon” if local power emission intensity is high and inefficient technologies, like resistive steam boilers or poorly flexibilized electrolytic hydrogen applications, replace existing systems. In large countries like China, relocating energy-intensive factories to regions with abundant renewable potential is an effective way to scale electrification with low-emission power[68].

Third, policies should be implemented now to hedge against the risk of a slower power sector transition. These "risk-hedging" technologies, such as CCU/S or CDR, should be deployed at an industrial scale by the end of this decade to ensure they can be scaled up in the 2030s[69,70]. REMIND scenarios with a fixed national budget show that more CDR will be needed to compensate for a delayed coal phase-out (Supplemental Material 9). The argument for CCU/S or CDR technologies is not that they are more cost-effective than renewables or electrification, but that due to uncertainty in coal phase-out timing, China must invest heavily in these technologies this decade. Therefore, near-term research should focus on feasibility and gathering the data needed for rapid deployment, such as provincial CCS potential and cost curves, to support timely infrastructure planning and siting[71–73].

This study designs coal power pathways within China's regulatory context and political constraints, but the model has limitations in fully capturing the complexity of regulations or their interactions with other sectors, e.g. end-use electrification or financial implications of coal phase-out. For example, near-term regulatory concerns may drive oversized battery storage to make renewables seem more “firm”, which would require high-resolution models to simulate accurately. Emission constraints could lead to large-scale battery deployment or new low-carbon dispatchable technologies, potentially raising electricity prices and discouraging electrification without further regulation. If power market reform allows for more flexible coal plant operations, it would also have financial implications, such as early retirements and debt write-offs. To ensure stability during electrification, maintaining coal plant financial health would require careful subsidy or capacity market calculations – too low or too high capacity payment could endanger the plan to rapidly electrify China's energy sector or to incentivize for more non-economic coal power plant build-out.

A key limitation of our analysis is the lack of spatial heterogeneity, as REMIND uses national averages rather than provincial data. This could introduce uncertainties, as regions like western China have abundant renewables, but EVs are mainly produced in the eastern provinces with high emissions. However, while trends like energy-intensive industries moving inland may reduce emissions faster than predicted, this is counterbalanced by slower infrastructure development for full supply-chain relocation. Additionally, emission intensities in Chinese provinces are more clustered than in European states, and therefore have relatively less spatial heterogeneity (see Figure S17). Therefore, despite the lack of spatial resolution, our main results should remain robust. Future work will incorporate higher-resolution power sector models into REMIND to improve regional analysis[74].

When it comes to phasing out coal power globally, there is no "one-size-fits-all" solution. Energy economists, grid operators, and climate scientists should help each country develop tailored climate solutions, considering local institutions, market structures, and pricing instruments[75]. As large-scale electrification has never occurred during coal phase-outs in industrialized nations (see Methods section), China's case provides valuable insights for countries with similar coal dependencies, political incentives for electrification, and strong economic growth prospects. These include large developing nations like India[76], South Africa[77] and Vietnam[78], which share similar concerns about energy import dependence and industrial upgrades.

# Methods

## Process-based IAM REMIND and its sector modeling methodology

The model used to assess China's mitigation pathway is REMIND v3.3.0.dev132 (REgional Model of INvestments and Development). It is a process-based IAM, linking global energy-economy-climate interactions[79]. REMIND has been frequently used in assessments of decarbonization scenarios, most notably in the IPCC[80–82]. The scenarios it produced have also helped make regional policies such as for the EU's 2030 goal ("fit for 55"), 2040 goal of emission reduction by 90%, as well as for Germany's energy transition plan. For each of the 12 regions, using a nested constant elasticity of substitution (CES) production function, the model maximizes interannual intertemporal welfare as a function of labor, capital, and energy use[79]. The macro-economic projections of REMIND come from various established global socio-economic scenarios jointly used by social scientists and economists – the so-called Shared Socioeconomic Pathways (SSPs)[83].

REMIND consists of several main modules connecting macro-economy, the energy system and climate. The energy system module is a linear conversion from primary energy to secondary energy, then to final energy. It is attached to the three energy demand modules transport, industry and buildings via final energy demand and prices. This coupling allows the model to incentivize or disincentivize technology investment through pricing signals. Under the climate mitigation scenarios, we set emission budgets for the globe or for the region. An endogenous $CO_2$ price is generated from the emission constraints, which is then added to the final energy price. In this way, the $CO_2$ pricing policy is transmitted to the demand sectors through raised prices of emission intensive fuels. However, in each demand sector, mitigation policies can also respond to subsidies, taxes as well as industrial policies such as infrastructure build-out of charging stations which facilitate the EV adoption.

## Transport module

In REMIND, the transport sector module "Energy demand GEnerator - Transport" ("EDGE-T") is bidirectionally soft-linked to the core energy system model, where the diffusion of transport technologies are modeled explicitly using logit regression functions in the detailed transport model[84]. In this sub-model, the rates at which the technologies diffuse are co-determined by sectoral policies as well as energy prices (including endogenous $CO_2$ prices) determined from the core energy system module in REMIND. This way of modeling reflects the reality that in practice, the rates of technological diffusion in the transport sector are mostly determined by specific sectoral policies and less so by competition between energy prices. This is suitable for the case of China, where the decisions to develop EVs are partially motivated by factors other than costs (such as oil import dependence).

The logit-based transport model EDGE-T does not include price-demand elasticity, whereas REMIND does. In EDGE-T, the market share of a technology in a country at a given time is determined by a logit function, where the inconvenient costs, prices of fuel, non-fuel costs, and the value of time together affect the outcome. The diffusion rate of a technology mostly depends on subsidies, investment costs, and inconvenience costs (for EV, this encodes information about charging infrastructure, EV battery range, and charging time). Electricity prices therefore play a small role in the logit-based transport scenario. Price-dependent elasticities also play a smaller role in the CES function in REMIND, as time-dependent elasticities are assumed to be lower in the 2030s compared to the long term. This formulation reflects the real-world dynamics that upfront costs, such as investment in electric technologies and lack of charging infrastructure, are major barriers to electrification.

In the transport sector module, there are overall four levels of policy ambition. The most pessimistic scenario "Mix1" is consistent with a no- to low-mitigation scenario: low BEV/FCEV ("fuel cell electric vehicle") vehicle shares among light-duty-vehicles (LDVs), corresponding to an electrification rate of 25-35% for LDV energy service demands in 2050. Similar vehicle shares are assumed for trucks. There is, in addition, very little mode shifting (under SSP2 assumptions). On the other extreme, the most optimistic policy setting "Mix4" is consistent with a global 1.5°C or well-below 2°C target. Under this setting, BEV shares in LDVs are high, and the electrification rates for LDVs are almost 100% in 2050 across regions. Trucks and buses have high electrification shares of up to 80% and 70% in 2050 respectively. Electric train shares go up to 85% globally (from 60% in 2015). Modal shares are not changed for developing regions, and moderately for developed regions. There are in addition two intermediate scenarios "Mix2" and "Mix3", where about 1/3 or 2/3 of the ambition of the most ambitious policy scenario is achieved. For all scenarios, aviation sees little hydrogen fuel in 2050. Ships currently have no electrification possibilities in the module, but can choose alternative liquid fuel use such as biofuel or synthetic fuel based on electrolytic hydrogen.

## Industry module

The industry module of REMIND v3.3.0.dev132 is built on the internal nested CES tree of the industry subsector (see prior version v3.1.0[47]), consisting of four nodes "cement", "steel", "chemicals" and "other industries", except in the newer version the "steel" CES branch is replaced with a process-based steel production model (see Figure S19). A set of key steel-making technologies are each represented by their capital and operational costs, specific energy and material inputs, and emissions. This provides an improved depiction of transformation dynamics and more detailed information on the choice of process routes and associated cost components. The main four implemented technologies are coupled in a two-stage configuration as shown in Figure 6. Technologies can switch between different operation modes with different inputs. Emitting technologies can optionally be equipped with carbon capture technology. Together, this results in six possible steel production routes, as shown in Figure S10. Such an industrial-process based modeling improvement as in steel is yet to be applied to the other three nodes "cement", "chemicals" and "other industries". In general, unless specifically assumed, the industry demand is generated via a model "Energy demand GEnerator - Industry" ("EDGE-Industry").

To project China's steel production into the future, we made an exogenous assumption based on a slow gradual linear decline of national production from current level to 70% around 2050, which is consistent with the trend of real estate slump. (Cement has a similar demand projection.) Cement has a similar demand projection, decreasing linearly from current level 2020 to 70% – namely around 1500 Mt/yr in 2050. There is currently no endogenous steel trade implemented in the model, historical trade patterns simply persist.

## Building module

The building sector's energy demand is projected in a simulation-based model "Energy demand GEnerator - Building" ("EDGE-B"), which divides the buildings sector demand into five main end uses: space heating, water heating, space cooling, appliances and lighting, and cooking. The model projects by way of simulation future energy demand based on historical trends, with GDP per capita being a key driver for most end uses. For space heating and cooling, demand is also influenced by climatic factors, which is captured by degree days. Space heating and cooling increase with floor space, which grows with affluence, though improvements in building insulation partly offset this effect. Water heating and cooking are assumed to reach saturation at high levels of affluence, while appliance and lighting use continues to rise with GDP per capita, although the elasticity weakens at higher levels. Technical efficiency improvements, especially in space cooling and heat pumps, reduce the energy demand for these end uses.

In the baseline ("current policy" / "NPi") scenario, electrification trends are driven by the dynamics of energy demand, with electricity shares growing in response to the high demand for electric end uses like appliances, lighting, and space cooling. The non-linear adoption of air conditioning, especially above certain GDP thresholds, contributes significantly to the increase in electricity consumption. At the same time, efficiency gains help balance the growth in electricity demand. The final energy mix continues to evolve, with coal and traditional biomass use declining, though policy interventions aimed at improving air quality are not explicitly modeled.

Under scenarios where climate constraints and policies are present, energy carriers' relative prices shift, leading to further electrification as fossil fuels become more expensive and the electricity supply is decarbonized. To accelerate electrification efforts, further financial incentives for technologies like heat pumps are implemented, aligning with the broader transition strategy of combining electrification with greening the power sector.

## Power sector module

The electricity sector in REMIND is parameterized using hourly model results based on region-specific load curves and renewable potential. This result is obtained using the DIMES model ("Dispatch and Investment Model for Electricity Storage"), a one-node hourly model with a 1-year (8760 hours) horizon that optimizes power sector investment and dispatch, which had been validated with another spatially heterogeneous power sector model. The DIMES model requires three time series for each region: load, wind power, and solar PV power, all with hourly resolution. The model minimizes total power supply costs, including investment, operation, maintenance, and fuel costs, by optimizing both investment (using a green-field approach) and the dispatch of non-VRE power capacity, along with short-term storage and reservoir sizes. Investments are calculated using annualized costs.

The DIMES version used to parametrize the REMIND-China power sector has been validated using the multi-nodal REMix model, which explicitly considers spatial heterogeneity across 15 European countries[85]. Recent high-resolution studies of the Chinese power grid typically use one node per province[37–39], analogous to the country-level resolution used in REMix for Europe. Therefore, using DIMES – which has been validated and improved based on REMix, is sufficient for analyzing the aggregate emission levels of the power sector and the broader economy in this study.

In order to parametrize a temporally complex future power system, a large parameter space needs to be explored. Due to computational constraints required in sweeping through such a large space, a computationally light-weight one-node model DIMES is used. REMix is only indirectly used in a validation step to check the result of the DIMES runs and to iteratively improve DIMES' representations on storage (see Figure S20). REMix's geographical coverage is limited to Europe, and therefore is only used to validate DIMES' approaches, not to directly parametrize REMIND. When comparing REMix and DIMES results for storage requirements and residual load duration curves under different wind and solar shares, DIMES was slightly adjusted to align with REMix results for Europe by splitting short-term storage into two technologies with costs varied by ±20%. This modified DIMES version is then used for the 11 regions in global REMIND, where each region's specific load and weather data is fed into DIMES. The results are incorporated into REMIND through the parametrization of the residual load duration curve[83].

Electrolyzers in REMIND are grid-based, but assumed to be flexible, i.e. they only produce at periods of low cost of electricity, i.e. when renewables are producing to abundance. This flexibility is modeled using data from the power system model Enertile for high variable renewable scenarios of Germany[87]. We use this study to derive a capacity factor for the electrolyzers, as well as the relation between the share of electrolysis in annual electricity demand, and the average electricity price for electrolysis normalized by the annual average electricity price. In this model, electrolysis can operate flexibly during renewable surpluses, as seen in recent Chinese projects, where hydrogen production emits no $CO_2$[58–62].

## LCA module

The LCA module is linked to the REMIND results in a post-processing step. The life-cycle carbon footprint for the LDVs (Figure 4C and 4F) are calculated using the open source software brightway2 (version 2.4.3)[88]. It is based on the prospective LCA (pLCA) databases created by premise (version 2.1.5)[116] and on the source LCA database ecoinvent 3.10[89]. Each pLCA database is aligned with a given IAM scenario in a given year (i.e. one database for each scenario and year), by integrating the IAM scenario data and additional life cycle inventories into the database. A detailed diagram illustrating the workflow of LCA and IAM integration is shown in Figure S21.

For our assessment of future life-cycle impacts of transport activities, the following steps of IAM integration are central (a full list of steps in the creation of pLCA databases can be accessed here: https://premise.readthedocs.io/en/latest/):

1. Efficiency Adjustments: Efficiency improvements in energy generation technologies (e.g., solar panels or electrolyzers), as provided by the REMIND module, are used to scale life-cycle inventories. More efficient solar panels, for example, reduce the upstream production and material extraction impacts, i.e. the "indirect emission" of electricity production.
2. Emission Factor Adjustments: Emission factors for power generation technologies are adapted based on the REMIND model. For instance, if coal power generation in China becomes less emission-intensive in a given scenario, this reduction is reflected in the LCA.
3. Market Shares: Output data from the REMIND model are used to build regional energy markets, such as the electricity and hydrogen markets in China. This influences the "direct emission intensity" of electricity or hydrogen.

In pLCA database, by explicitly building a full supply chain for China's vehicle production and usage, we consider both the "indirect" and "direct" emission of electricity production, i.e. the carbon footprint of electricity, when calculating the LCA carbon footprint impact of transport activities in China (see Figure S21).

Our assessment compares battery electric and gasoline engine passenger cars of medium size by year. In 2020, the BEV dataset represents a vehicle with a 141 kW electric motor, a curb mass of 1764 kg, of which 400 kg is the battery mass, and a full battery range of 196 km. The gasoline car data represents a vehicle with a 126 kW combustion engine, a curb mass of 1579 kg, and a full tank range of 1002 km. Beyond 2020, the datasets chosen as functional units of the LCA represent fleet averages in the given (scenario) year. GHG metrics from the IPCC AR6 report are used as characterization factors in the life cycle impact assessment[90].

## Global, regional and sectoral scenario design and assumptions

The mitigation scenarios studied here are based on several constraints and assumptions, both globally, for specific regions, and for specific sectors. First, under the global settings, we use a SSP2 "middle-of-the-road" scenario[83], which projects population and GDP globally for each region under existing trends, and are exogenously set in the model. In addition, we constrain the global $CO_2$ emission budget (from 2020 to 2100) to 1150 Gt, which is compatible with the global budget for the climate target of 67% probability of limiting warming to 2°C in AR6[91] (we used AR6 and not the newest update on carbon budget[92] to stay consistent with the other parts of the IAM, but extrapolated until 2023 to obtain the remaining carbon budget used in the post-modeling analysis on cumulative emission). Second, for high current emission regions, we set $CO_2$ net-zero annual emission target-year for China (2060), Europe (2050), U.S. (2050), and India (2070) according to their respective climate pledges.

In the climate constrained scenarios, we are using high electrification scenario assumptions, which is based on the current trend of rapidly declining renewable costs[40], and is in line with many international and domestic modeling results of China under a net-zero pathway[43,50,93]. The high direct electrification scenario aligns with the Chinese energy policy trend on electrification, and is a very attractive option due to its energy-efficiency potentials in industry[94]. In addition, in the context of aggravating water scarcity under rapid societal developmental changes as well as climate impact[95], a high direct or indirect electrification instead of a high biomass scenario is considered appropriate for China. The modeling of the Chinese transport sector in REMIND has taken into account the latest sales data until the end of 2022 (see Figure S7).

In terms of climate and sectoral policies, we emphasize the two dimensionality of the scenario design (see Table 1): electrification rates and coal phase-out speeds. Electrification in our model is not entirely exogenously determined (i.e. set a-priori, see Table 2). It partially results from climate constraints and sectoral policies. We model demand-side electrification alongside supply-side coal phase-out scenarios to examine their interaction and assess emissions impacts.

In this study, there are several implicit assumptions on coal-based power generation technologies. First, in REMIND, when there are sufficient constraints on global emission (such as Paris aligned scenarios), coal power plants are never profitable in China in the model REMIND. Therefore the optimal path for the model to mitigate coal power emission is to always phase it out as quickly as possible and phase in the more economic options such as wind and solar. However, due to political economy and operational factors exogenous to power market fundamentals, an extra parameter is implemented which designates the upper limit by which coal capacity can be phased out in a year. This parameter, together with the hard-coded coal plant time-dependent capacity factors estimated from bottom-up results, is adjusted to produce the various national coal phase-out trajectories (see Table S1 and S2).

Second, due to the fact that coal retrofit CCS is not practical or economical in a world where coal power plants will mostly have generation peaking roles in the system, we do not have it as a technological option in the model. First, coal retrofit CCS is technically difficult to do in practice, since retrofitting usually requires a redesign of the layout to create room for the extra equipment to be put in, as well as nearby $CO_2$ storage site[96–98]. Second, assuming eventually renewable power will reach majority share, implying a “peaking” role for the coal power plants, deploying CCS on a declining capacity factor makes little economic sense. So essentially the two practical and economic options for coal plants would be to either to let coal plants lower their capacity factor to accommodate cheaper and greener renewables, or to have new-build coal CCS plants (this is in line also with the US regulation on coal plant retirement[99]). In terms of new-built coal power plants with CCS, there are only two combined-heat-and-power plants in operation (0.65 $MtCO_2$/yr capture volume) and three under construction in China (2.7 $MtCO_2$/yr capture volume) to be operational by 2025[100]. There are currently no coal retrofit CCS plants in China or in the world. Therefore we explicitly include only new-built coal power plants, and exclude retrofit coal power plants as a technology option. If CCS retrofit or new built were to be part of the solution, not only do they need to have high capture rate to minimize requirement for negative emission compensation, which implies higher cost[101], they would also have to scale up from almost zero to at least the current capacity of coal power plant, which is 1170 GW in 2024[102]. The scale-up rate required would be similar to the speed at which unabated coal power is phased out in our model by 2050, implying around 40 GW/yr on average, or at least 17 pp/yr growth rate for 30 years, starting at 10 GW in the base year. Besides the obvious scale-up problem, the availability of carbon transport and storage is also constrained in the near term.

## Coal power phase-out scenario design based on feasibility concerns

Informed by recent literature on the feasibility of coal phase-out in non-OECD countries in IAM scenarios[44–46], when we design the scenarios for China, we pay special attention to the feasibility of rapid coal power phase-out. Currently, the share of coal power generation in China remains high at around 61% in 2022, even though it has been on a decreasing trend in the last two decades. Given the rising shares of renewable energy and record-breaking renewable additions, capacity factors of coal plants will certainly be lowered despite new coal power capacity additions[41,103,104]. However, how fast unabated coal power can be realistically phased out is an open question.

In the “fast” scenario, a full phase-out of coal power is projected by 2040, with its share decreasing by an average of 4.1 pp/year from 2022 to 2037. The decline peaks at 6.1 pp/yr in 2030. This rate is 15% higher than the highest rate which was achieved in economically advanced countries during brief periods lasting several years, typically without simultaneous end-use electrification (e.g., the UK with 5.3 pp/yr, as discussed below). Nevertheless, the “fast” scenario reflects common coal phase-out pathways in China under 2°C or 1.5°C scenarios in the IPCC AR6. Considering the difficulties of coal power retro-fit mentioned above, this rapid coal phase-out scenario poses serious challenges to the adaptivity of China’s power system. In contrast, for the “plateau30” scenario, the share decline rate is around 2.8 pp/yr in the 2030s, until full phase-out in 30 years. Its coal share declines at 2.2 pp/yr on average from 2020 to 2060, which matches the U.S. – a larger, a more geographically heterogenous country than Germany and closer to China, during its fastest coal phase-down period in 2008–2018 also with coal-to-gas switch and little end-use electrification. The “medium” path decline rate of 3.5 pp/yr in the 2030s matches the fast coal phase-down period between 2015 and 2020 in Germany under partial coal-to-gas switch and almost no additional sectoral electrification. By this way of cross comparison, the most “conservative” scenario is “NPi”, with 2.3 pp/yr in the late 2020s and 2030s, and an average of 1.6 pp/yr between 2020 and 2060, eventually phasing out coal power in 2060.

Despite such rough comparison as well as plenty of literature drawing the historical comparison of coal phase-out speeds in various countries such as the U.S., UK and Germany to project the transitions of developing countries, this metric alone cannot tell us very much about the case in China. First, in China’s net-zero path, there is a simultaneous rapid electrification push, which did not occur in historical episodes in Europe and the U.S. For example, Germany has 44% share of wind and solar but only 20.2% electrification rate in 2022, whereas China has 11% variable renewable share but as high as 27% electrification rate in 2021, and has ambition to reach 30% by 2025[105]. Overall, the power demand will drastically increase in China (roughly 2-fold over 40 years) while the grid is being decarbonized (Supplemental Material 2). Second, past coal-phase out episodes were accomplished with a large amount of transient coal-to-gas switch (the so-called “gas bridge”). For example, since 1991 in about three decades the UK has almost completely phased out coal in the power sector, with around 1% coal power share in 2022. During this time, there were two distinct rapid coal phase-out phases, when the coal power share change was around -4.5 pp/yr from 1991 to 1999, and -5.3 pp/yr from 2012 to 2019. However, after coal phase-out it still has a relatively large fossil fuel share in power generation (40.8% in 2022), consisting almost entirely of gas[106]. In the U.S. the fossil fuel share in power generation is even higher (60.2%)[107]. Coal and gas power are both dispatchable forms of generation, therefore in terms of dispatch and transmission operation, the two sources are similar enough that a switch could happen very quickly. However, a “gas bridge” is commonly not seen as an economic option for China in the near- or medium-term[41,108], nor is it long-term climate compatible considering most major emitters must go to near-zero emissions around the middle of the century[109].

Despite that the Chinese power sector transition may face unprecedented challenges in achieving rapid speed, using the historical rates from ten to several years ago in other countries as a reference for China could also be overly conservative. In the past decade, the cost of onshore wind and solar power in China has decreased by 30% and 75%. China has cheap renewable potentials, and the business environment for renewables is different from ten years ago. Additionally, institutional and operational learnings have also made renewable integration more mature, policy proposal around renewable integration is being discussed[110,111], transmission and storage has also grown and increasingly strengthened[112,113]. Deployment of renewables are increasingly seen as measures which increase food and energy security and sovereignty against global energy and food price inflation or volatility. In light of these nuances in assessing feasibility pathways, we chose "NPi" and "fast" coal power phase-out scenarios as extreme cases: the former represents a reasonably optimistic path based on historical cases as well as domestic results, the latter represents AR6 pathways which are seen as overly optimistic in the literature[44,45].

Despite the currently fragmented power markets as well as decentralized regional governance, the Chinese system of new energy production has proved itself to be resilient against short-term supply-chain shocks. Therefore, in this study we only consider the sensitivity which has to do with 5 to 15 years of coal power phase-out delay which could be due to semi-persistent, structural drivers which hinder the adoption of low carbon power generation. An example of such system resilience to shocks is that in 2021 China added record capacity of solar generation (53 GW), despite the solar-grade polysilicon price spike by 300% due to the power shortage[114]. Due to increased supply, the price eventually comes down to the normal level before. Another example was the perceived "security of coal", but due to awareness of climate protection, this is also increasingly weakened as energy security is also increasingly linked to renewables in official policies[115]. Therefore, delays even longer than 10 years can be considered a low probability event, which is outside the scope of this analysis.

# Resource availability

## Lead Contact

Requests for further information and resources should be directed to and will be fulfilled by the lead contact, Chen Chris Gong (chen.gong@pik-potsdam.de).

## Materials Availability

This study did not generate new unique reagents.

## Data and Code Availability

1. Data
   Simulation results and plotting scripts are archived at: Zenodo
   https://zenodo.org/records/14895088
2. Code
   The model code with data updates and sectoral assumptions for China, including all input data apart from the IPCC scenarios, is available on GitHub:
   https://github.com/cchrisgong/remind/releases/tag/v3.3.0dev132_china. It is a release of the modified fork of the REMIND model v3.3.0.dev132
   https://github.com/remindmodel/remind/releases/tag/v3.3.0.dev132.

3. Additional information
   Any additional information required to reanalyze the data reported in this paper is available from the lead contact upon request.

# Acknowledgements

We are supported by the German Federal Ministry of Education and Research (BMBF) via the project INTEGRATE (FKZ 01LP1928A) and the project Kopernikus-Ariadne (FKZ 03SFK5N0, FKZ 03SFK5A). We thank Dr. Kejun Jiang, Dr. Shuwei Zhang, Felix Schreyer, Dr. Gabriel Abrahão, Prof. Shiyan Chang for discussion.

# Author contributions

C.C.G., F.U. and C.B. designed and coordinated the scenario development. Y.Y. and D.B. provided the life-cycle analysis results in the analysis. G.L., F.U. and C.B. performed theoretical and conceptual validation of the manuscript. C.C.G., R.P., J.H. and J.M.improved REMIND China's transport sector modeling to align with current development. J.D., M.P., S.M. and C.C.G improved REMIND China's industry sector modeling. R.H. helped validate REMIND China's building sector result. C.C.G prepared the manuscript with comments and contributions from all co-authors.

# Declaration of Interests

The authors declare no competing interests.

Supplemental Materials

## 1. Scenario designs for coal power phase-out pathways

For the power sector, under different coal phase-out scenarios, various time series of exogenous capacity factors as well as the maximum allowed early retirement of capacities are given as follows in order to produce slower coal power phase-out outcomes.

Table S1: Coal capacity and capacity factors used in four levels of coal phase-out pathways: "current policy" (NPi), "plateau30", "medium" and "fast", that are exogenously set in the scenarios. They represent four levels of ambitions of coal power phase-out timelines, all of which are consistent with the 2060 climate neutrality target. They are set in a wide range to indicate uncertainties due to political economy factors in the sector.

| Scenario | Variable | 2020 | 2025 | 2030 | 2035 | 2040 | 2045 | 2050 | 2055 | 2060 |
|---|---|---|---|---|---|---|---|---|---|---|
| **CP-NPi** | **coal capacity (GW)** | 1040 | 1130 | 1070 | 880 | 800 | 470 | 180 | 40 | 0 |
| | **average coal power plant capacity factor (%)** | 58 | 58 | 58 | 58 | 47 | 47 | 50 | 66 | - |

| **CP-plateau30** | **coal capacity (GW)** | 1040 | 990 | 930 | 790 | 600 | 350 | 80 | 0 | 0 |
|---|---|---|---|---|---|---|---|---|---|---|
| | **average coal power plant capacity factor (%)** | 58 | 63 | 63 | 48 | 35 | 22 | 10 | - | - |
| **CP-medium** | **coal capacity (GW)** | 1040 | 990 | 840 | 610 | 360 | 70 | 0 | 0 | 0 |
| | **average coal power plant capacity factor (%)** | 58 | 63 | 45 | 35 | 22 | 10 | - | - | - |
| **CP-fast** | **coal capacity (GW)** | 1040 | 1070 | 760 | 370 | 0 | 0 | 0 | 0 | 0 |
| | **average coal power plant capacity factor (%)** | 58 | 55 | 34 | 16 | - | - | - | - | - |

Table S2: Annual coal power generation, coal power share of total power generation and its decline rate in percentage point per year (pp/yr) for each 5-year period under the various coal power phase-out scenarios for China's net zero scenarios.

| **Scenario** | **Variable** | **2020** | **2025** | **2030** | **2035** | **2040** | **2045** | **2050** | **2055** | **2060** |
|---|---|---|---|---|---|---|---|---|---|---|
| **CP-NPi** | **coal power generation (TWh/yr)** | 5300 | 5720 | 5430 | 4460 | 3240 | 1930 | 760 | 160 | 8 |
| | **coal share (%)** | 66 | 55.6 | 44.0 | 31.8 | 21.6 | 12.8 | 5.3 | 1.1 | 0.1 |
| | **coal share decline rate (pp/yr)** | - | -2.1 | -2.3 | -2.4 | -2.0 | -1.8 | -1.5 | -0.8 | -0.2 |
| **CP-plateau30** | **coal power generation (TWh/yr)** | 5300 | 5500 | 5130 | 3280 | 1850 | 690 | 70 | 0 | 0 |
| | **coal share (%)** | 66 | 53.9 | 41.4 | 24.1 | 12.7 | 4.7 | 0.5 | 0 | 0 |
| | **coal share decline rate (pp/yr)** | - | -2.4 | -2.5 | -3.4 | -2.3 | -1.6 | -0.8 | -0.1 | 0 |
| **CP-medium** | **coal power generation (TWh/yr)** | 5300 | 5510 | 3300 | 1870 | 710 | 60 | 0 | 0 | 0 |
| | **coal share (%)** | 66 | 53.1 | 28.7 | 13.8 | 4.9 | 0.4 | 0 | 0 | 0 |
| | **coal share decline rate (pp/yr)** | - | -2.6 | -4.9 | -3.0 | -1.8 | -0.9 | -0.1 | 0 | 0 |
| **CP-fast** | **coal power generation (TWh/yr)** | 5300 | 5180 | 2230 | 530 | 0 | 0 | 0 | 0 | 0 |
| | **coal share (%)** | 66 | 51.2 | 20.2 | 4.2 | 0 | 0 | 0 | 0 | 0 |
| | **coal share decline rate (pp/yr)** | - | -3.0 | -6.1 | -3.2 | -0.8 | 0 | 0 | 0 | 0 |

## 2. Power supply and demand mix

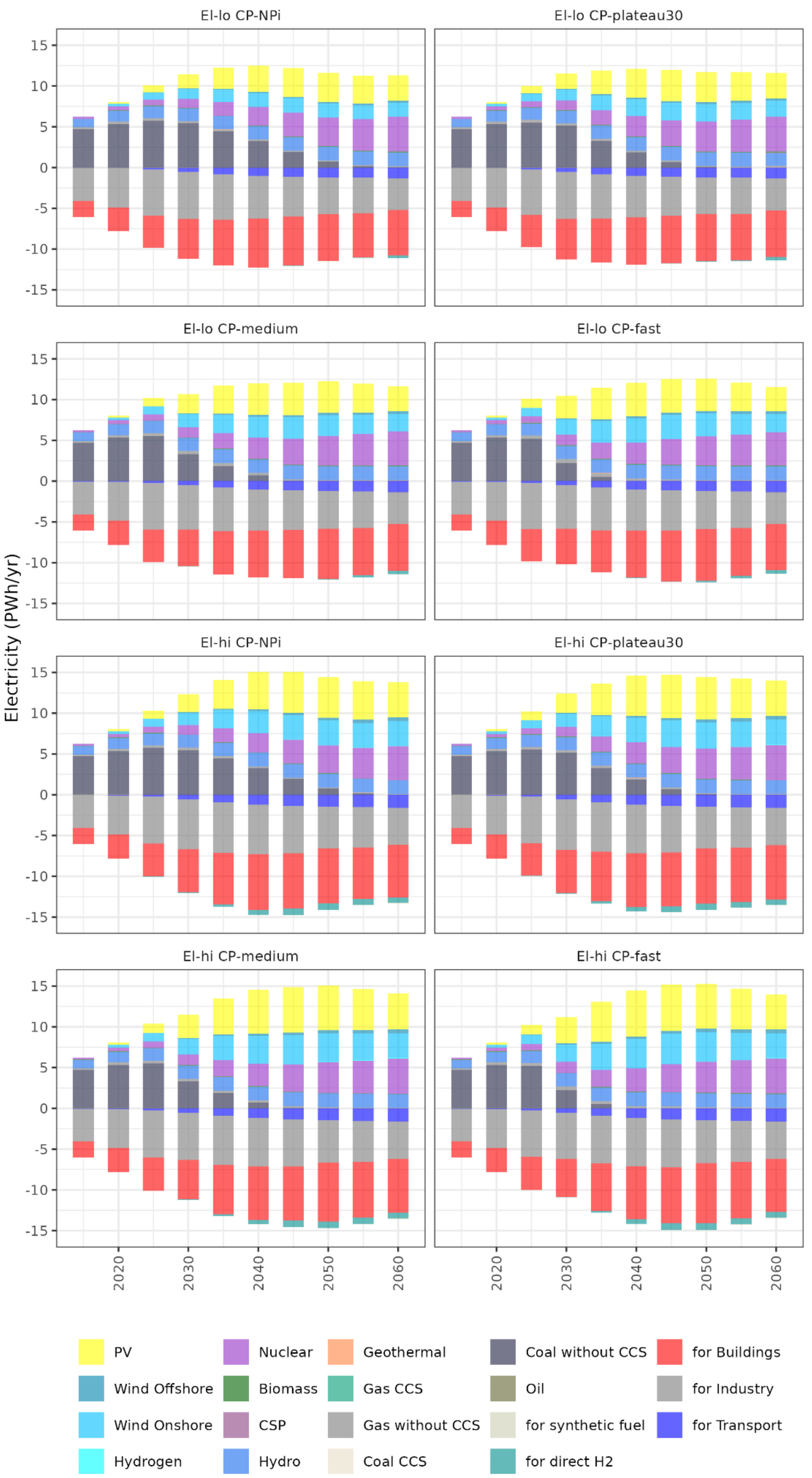

Figure S1: Mirroring power supply and demand mixes. We illustrate the demand of electricity on the lower axis for the ease of visualization.

Under mitigation scenarios (“El-hi”), the electricity demand sees an increase of almost three-fold due to the effect of electrification of end-use sectors. This is to compare to the roughly two fold increase under the relatively low ambition, low electrification scenarios. Under high electrification, due to heating and cooling increase, building sector power demand increases dramatically. Industry sector demand sees a moderate increase of around 30% in the near-term due to the increased share of secondary steel in the mix (see Supplemental Material 8 below), and falls again due to the moderate decrease of industry demand (due to population decrease under SSP2). Transport electrification contributes to around 10% of total electricity demand by 2060, whereas the production of hydrogen and synthetic fuel contributes to less than 10%. Most of the hydrogen in the mitigation scenario is supplied via biomass combined with carbon capture (around 60% to 80% across the scenarios), and only a small part (20-40%) is supplied by electrolytic hydrogen (not shown). The total electrolytic hydrogen in the scenario is around 17 million tonnes by 2050.

Under the fast coal power phase-out scenario, REMIND reacts to the higher intermediate electricity price (see Supplemental Material 4) by lowering the electrification rate, resulting in slightly less power demand in 2030. The rapid phase out of coal power in 2030 and 2035 is compensated by an increased rate of expansion of renewable and nuclear power compared to the slower phase out scenarios.

## 3. Primary energy mix

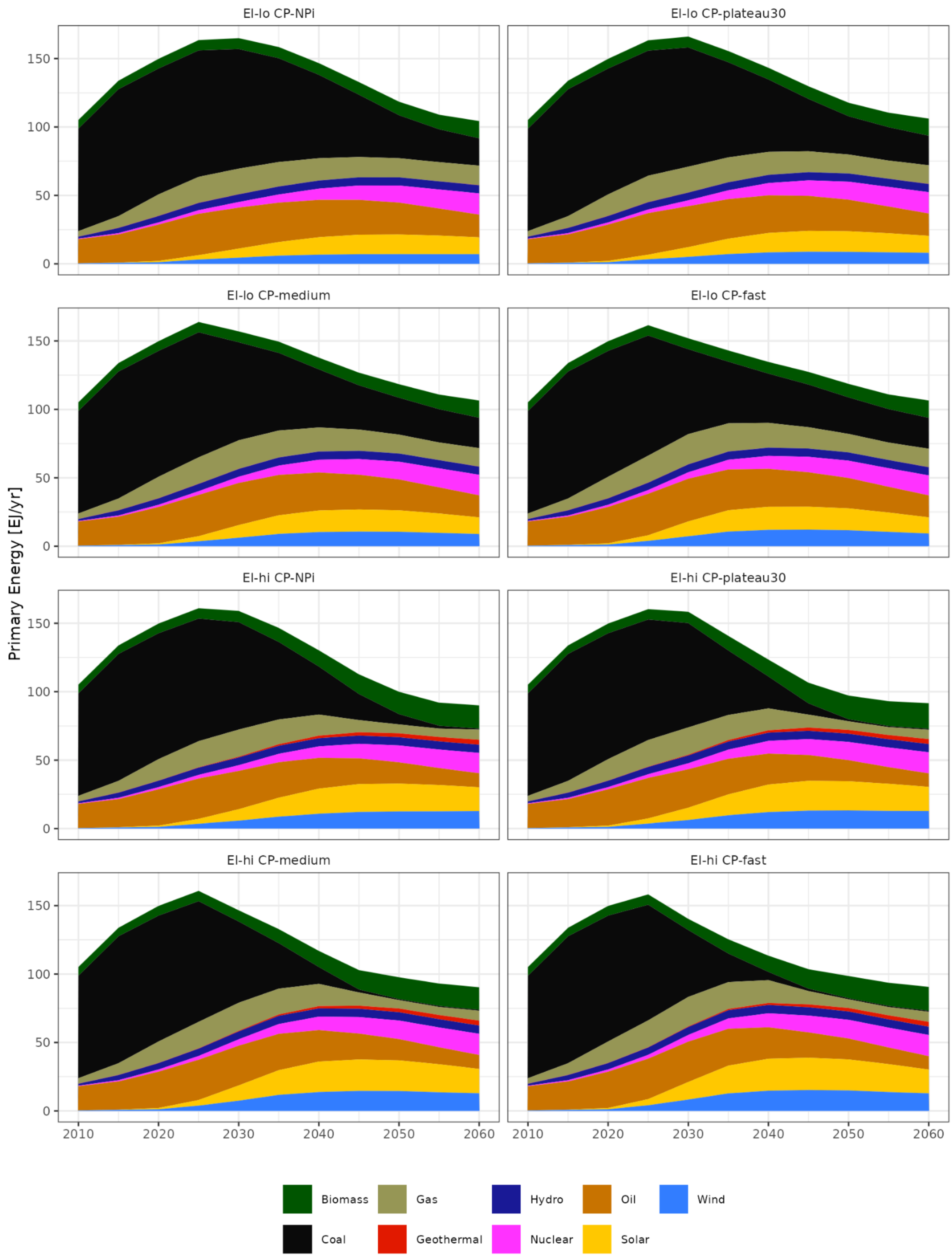

Figure S2: China's primary energy mix.

Even under low-ambition low-electrification reference scenarios ("El-lo"), coal, gas and oil shares decrease under existing trends, and make up around half of primary energy mix in 2060. Under climate policy scenarios ("El-hi"), an expansion of biomass, geothermal, hydro, nuclear, and renewables could replace a significant share of natural gas and oil (with gas around 7% and oil around 10% for "El-hi plateau30"), while coal is phased out.

4. Endogenous electricity and coal prices

(a)

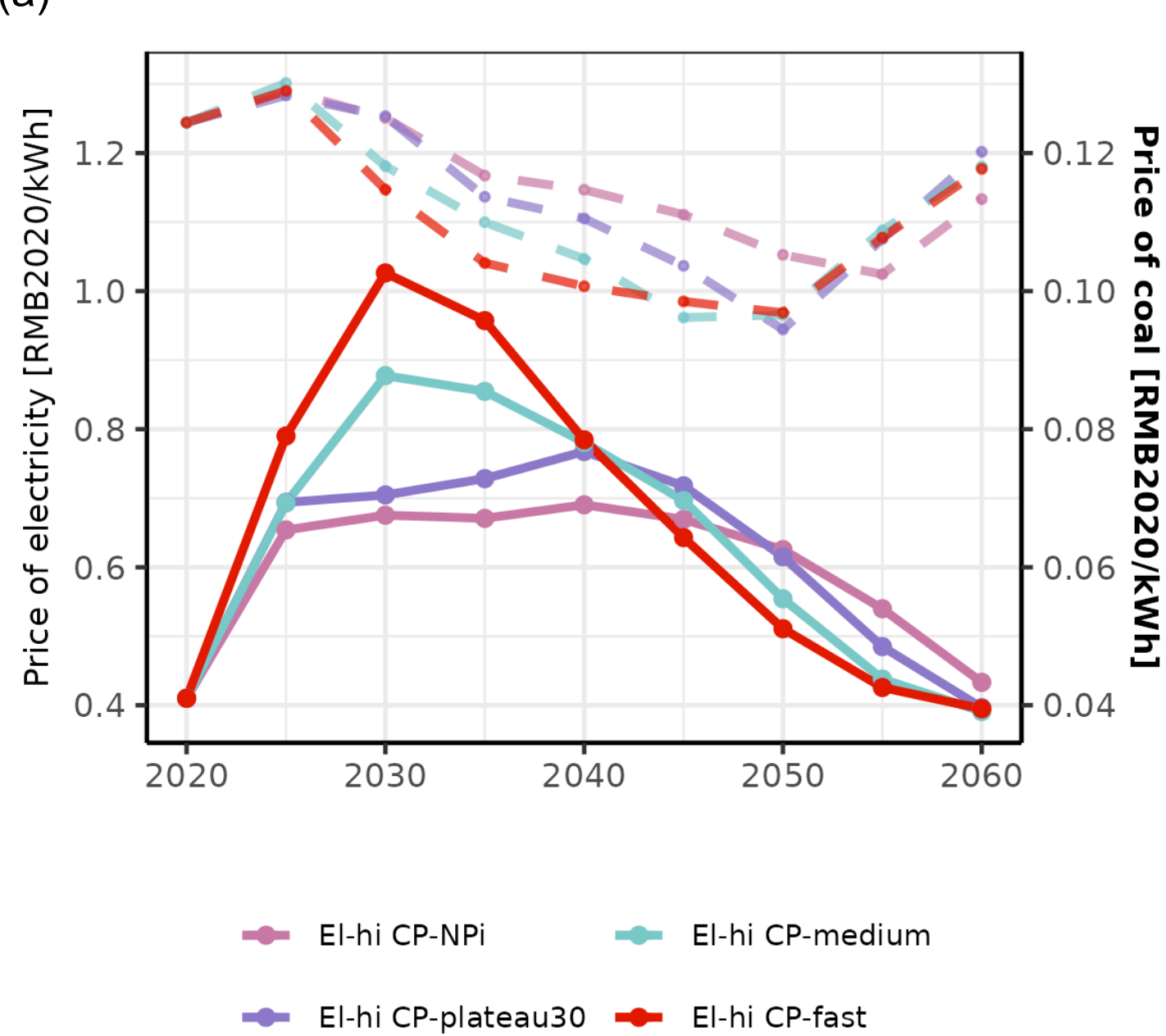

(b)

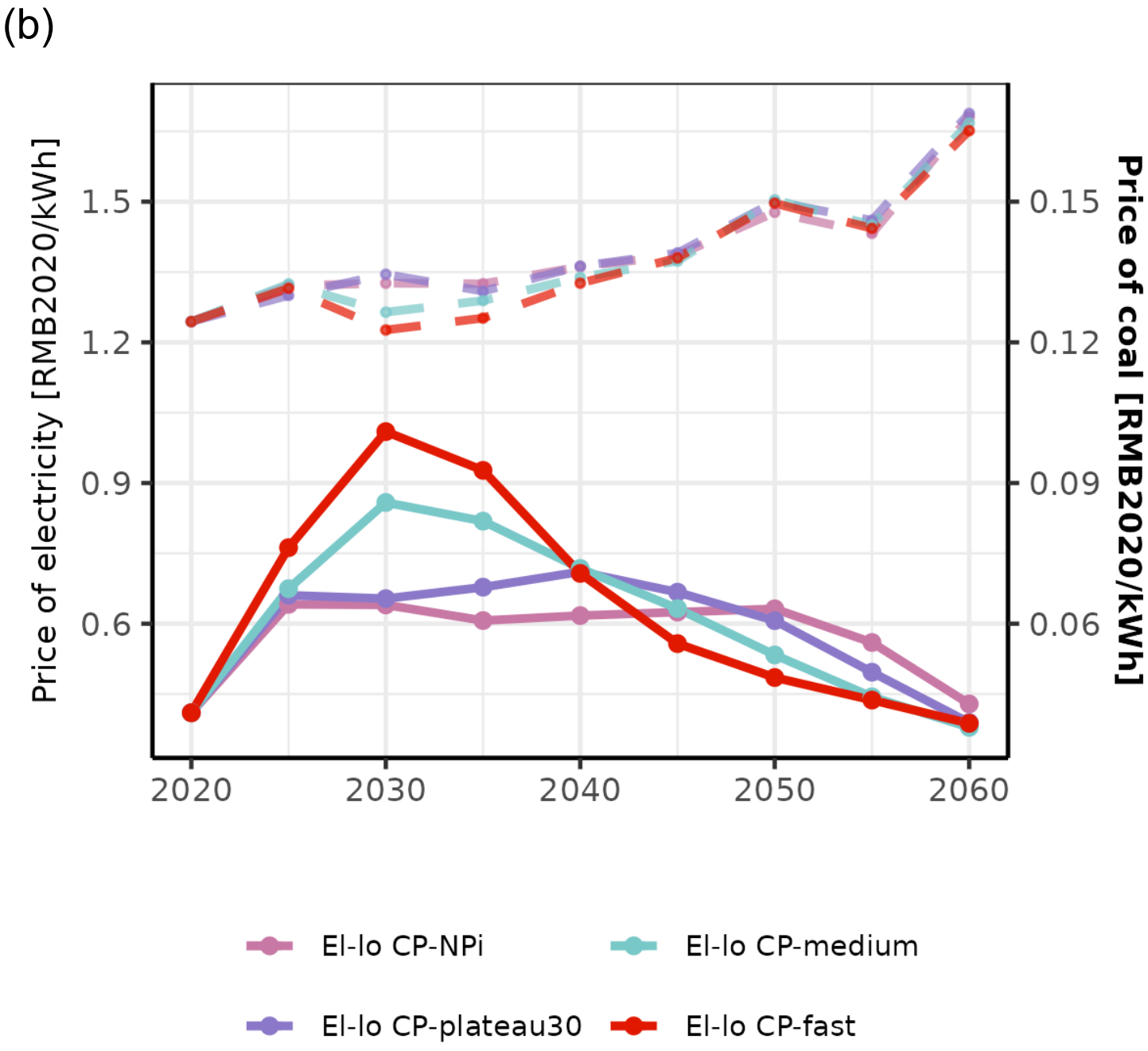

Figure S3: Annual average electricity price time series and coal price time series: (a) high-climate ambition, high-electrification mitigation scenario (“El-hi”); (b) low-ambition low-electrification reference scenarios (“El-lo”).

For mitigation scenarios, the electricity prices in the 2020s and 2030s are generally high due to the high capital expenditure of a massive investment of renewables and adjustment costs, i.e. the cost mark-ups for fast expansion of generation, grid and storage technologies. The faster coal power is phased out, the higher the near-term electricity price. But in the long-term, electricity prices are lower under faster phase-out scenarios due to the lower renewable power costs as a result of early technological learning and converge around 40 cent RMB/kWh.

Over the long-term, we find that under high electrification scenarios, the electricity price is 12% lower than in low electrification scenarios. Under the reference scenario, the electricity price converges to around 45 cent RMB/kWh in 2060, whereas in the mitigation scenario it converges to around 40 cent RMB/kWh. Due to a lack of electrification and hydrogen production, the overall lower electricity demand under reference scenarios means that renewables have a less steep learning curve, and therefore the electricity mix consisting mostly of wind and solar is more expensive under reference scenarios. However, the general tendency of electricity prices is not qualitatively different under mitigation constraints. This indicates that electricity price movement is largely determined by the energy mix, which undergoes a similar dramatic shift under the reference scenarios.

The model electricity price in 2024 is comparable to the current household electricity price of around 0.54 RMB/kWh[1]. It is only lower due to near-term fixing of generation quantities, which might overshoot power demand data, due to the fact that they are from different data sources. The primary coal price in the near-term is comparable to the current Chinese domestic coal price, which is around 0.12 to 0.2 RMB/kW[2]. The coal price is endogenous to the model[3,4].

As shown in Figure S3, coal prices in China are indirectly related to electricity prices but show strong anticorrelation. In our scenario, the faster coal is phased out, the faster coal prices drop. The primary energy coal price is determined by the coal demand and supply equation. Coal demand comes from sectors such as electricity, metallurgy (iron and steel, cement, aluminum), and rural heating. Coal is also used for liquid fuel production. The supply of coal in China is a combination of imported coal and domestic extraction. Coal is globally traded in REMIND using the Nash algorithm[5]. Extraction follows the Hotelling rule[3,4]. The marginal price in coal's demand and supply equation is the coal price in REMIND for a given period.

As coal use in electricity declines under mitigation scenarios, the price of coal also drops by around 20-30% in 2050 compared to 2025 (Figure S3(a)). Compared to the 2025 prices, coal price drops by around 24% under slow coal phase-out (with high electrification) by 2050, and drops by more than 30% under fast coal phase-out (with high electrification).

5. Selected results on emissions

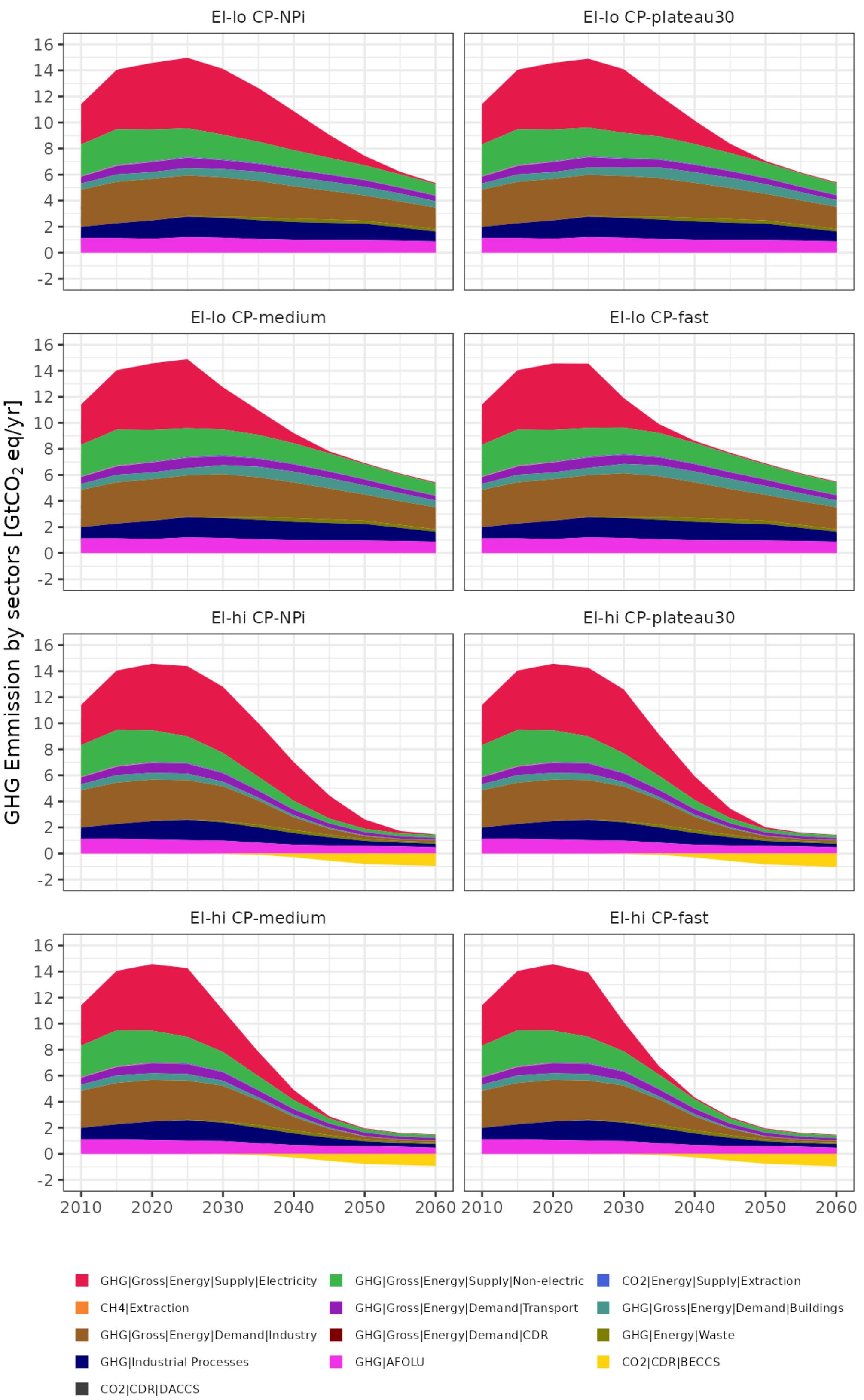

Figure S4: China's GHG annual emission by sector.

In low electrification reference scenarios (“El-lo”), only the power sector undergoes transformation, resulting in limited abatement in the demand sectors and non-electric supplies. To achieve net-zero emissions, demand sectors should be electrified as directly as possible, as this is the most cost-effective option compared to indirect electrification alternatives, such as hydrogen.

As shown in Figure S4, under the mitigation scenarios constrained by the 2060 net-zero $CO_2$ pledge (“El-hi”), China’s total greenhouse gas (GHG) emissions will not reach zero but will remain slightly positive. Take “El-hi CP-plateau30” scenario, in 2060, the residual GHG emissions primarily come from Agriculture, Forestry, and Other Land Use (AFOLU) (478 $MtCO_2$/yr), industrial processes (265 $MtCO_2$/yr), and some energy-related emissions (Industry 150 $MtCO_2$/yr, Waste 120 $MtCO_2$ /yr, Transport 147 $MtCO_2$/yr and non-electric energy supply 204 $MtCO_2$/yr). These are offset by more than 1 $GtCO_2$/yr of negative emissions through Bioenergy with Carbon Capture and Storage (BECCS) in most mitigation scenarios. Direct Air Capture with Carbon Capture and Storage (DACCS) is not a significant net-negative emission option due to its higher cost compared to BECCS.

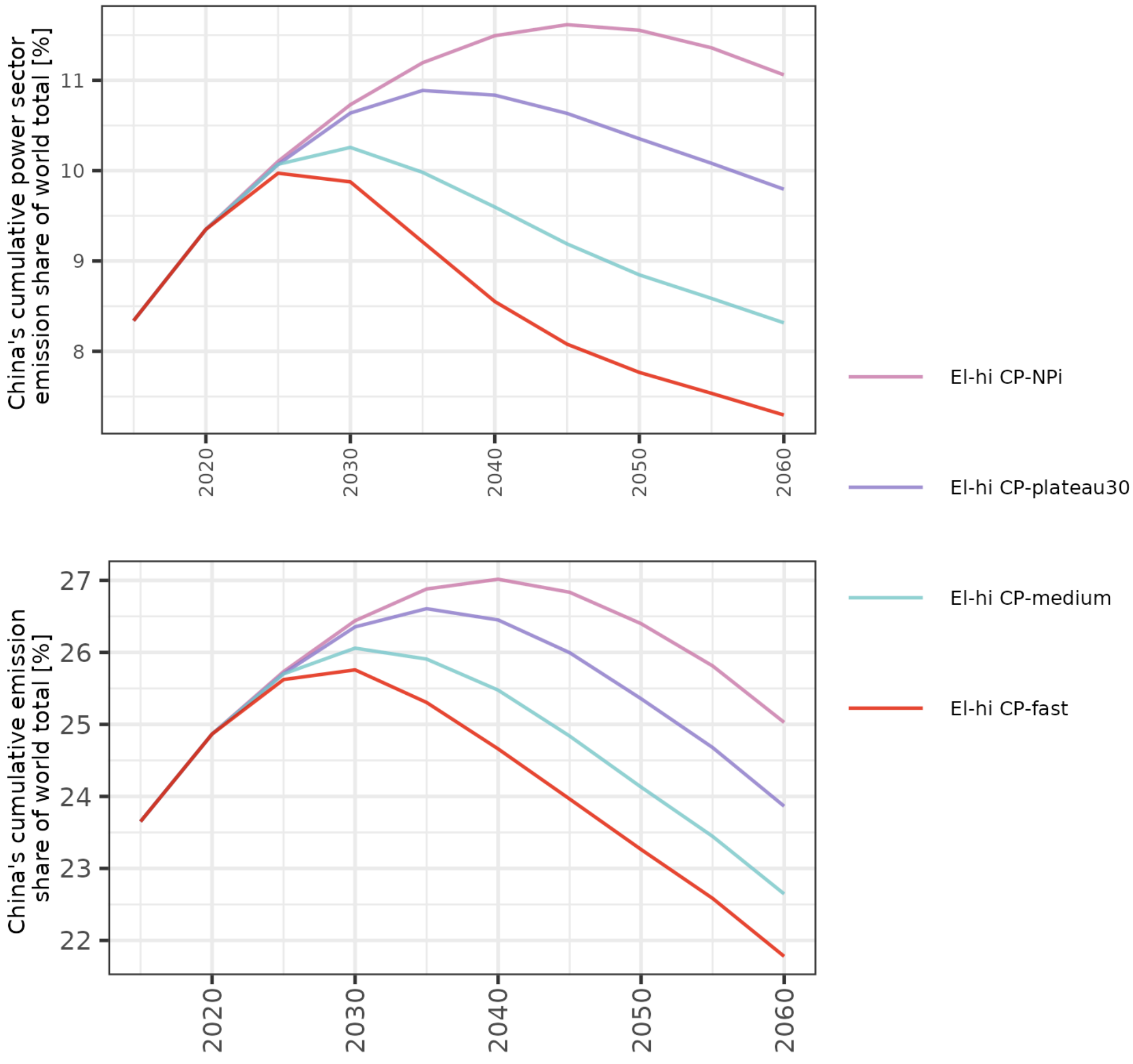

Figure S5: China’s cumulative emission as a share of the world’s economy-wide cumulative emission (since 2005). Top: China’s cumulative emission from the power sector alone as a share of the total world cumulative emission as a function of time under various mitigation scenarios (“El-hi”). Bottom: China’s economy-wide cumulative emission as a share of total world cumulative emission.

Under the mitigation scenario, major emitters such as China, the U.S., and the EU all reach net-zero emissions around the middle of the century. Since 2005, China's total contribution to global warming is projected to be around 22-25.4% until 2020. Then later, depending on the pace of the coal phase-out, it could be as high as 27% under NPi or 24.5% under “fast”, in 2040. Notably, as Figure S5 shows, not only does China's power sector account for a significant share of annual global emissions in 2020, but the cumulative emissions from this single sector in one region also represent a non-trivial portion of global economy-wide cumulative emissions, potentially reaching more than 11.5% if unabated coal power is only completely phased out after 2055. This contrasts with a more optimistic scenario of a rapid coal phase-out by 2040, where the contribution would be around 10%.

6. Buildings sector energy mix

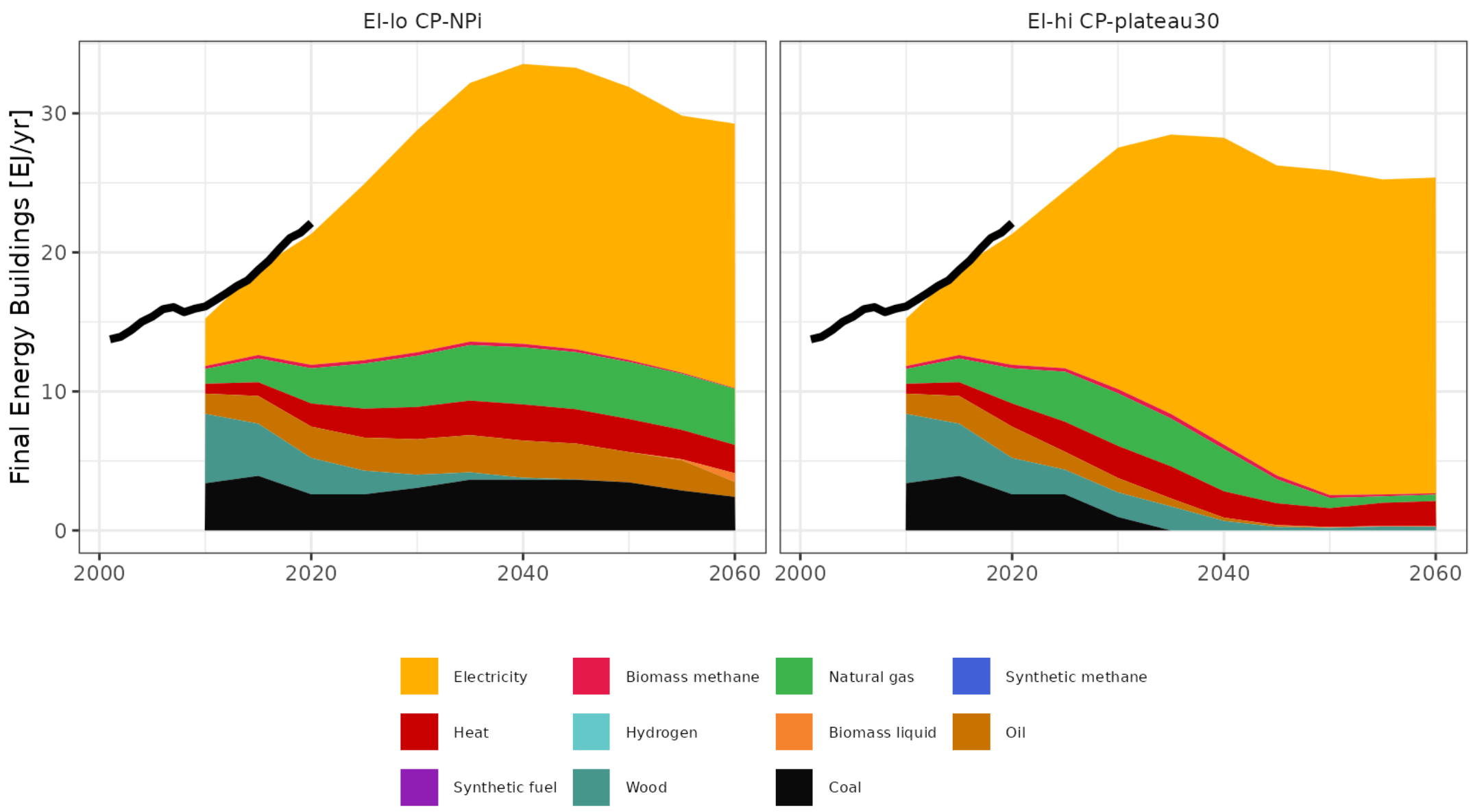

Figure S6: China’s building sector final energy mix comparing a representative reference policy scenario “El-lo CP-NPi” and a representative climate scenario “El-hi CP-plateau30”.

As shown in Figure S6, final energy demand increases as the population becomes more affluent, leading to greater use of cooling services and more electrified household appliances. To meet this growing demand while replacing mostly oil and coal under a climate scenario, there is an expansion of electric heat pumps, natural gas, and district heating in the late 2020s and 2030s. Additionally, wood use is partially replaced as the population becomes more urbanized and people seek to reduce indoor air pollution. By 2060, almost all heating services are provided through either district heating or heat pumps. Wood is continuously used until 2045, while hydrogen is not used at all for heating due to the existence of cheaper options for direct electrification. Gas is used as a transition fuel from 2020 until 2045, when it is finally replaced by heat pumps.

7. Selected transport sector results

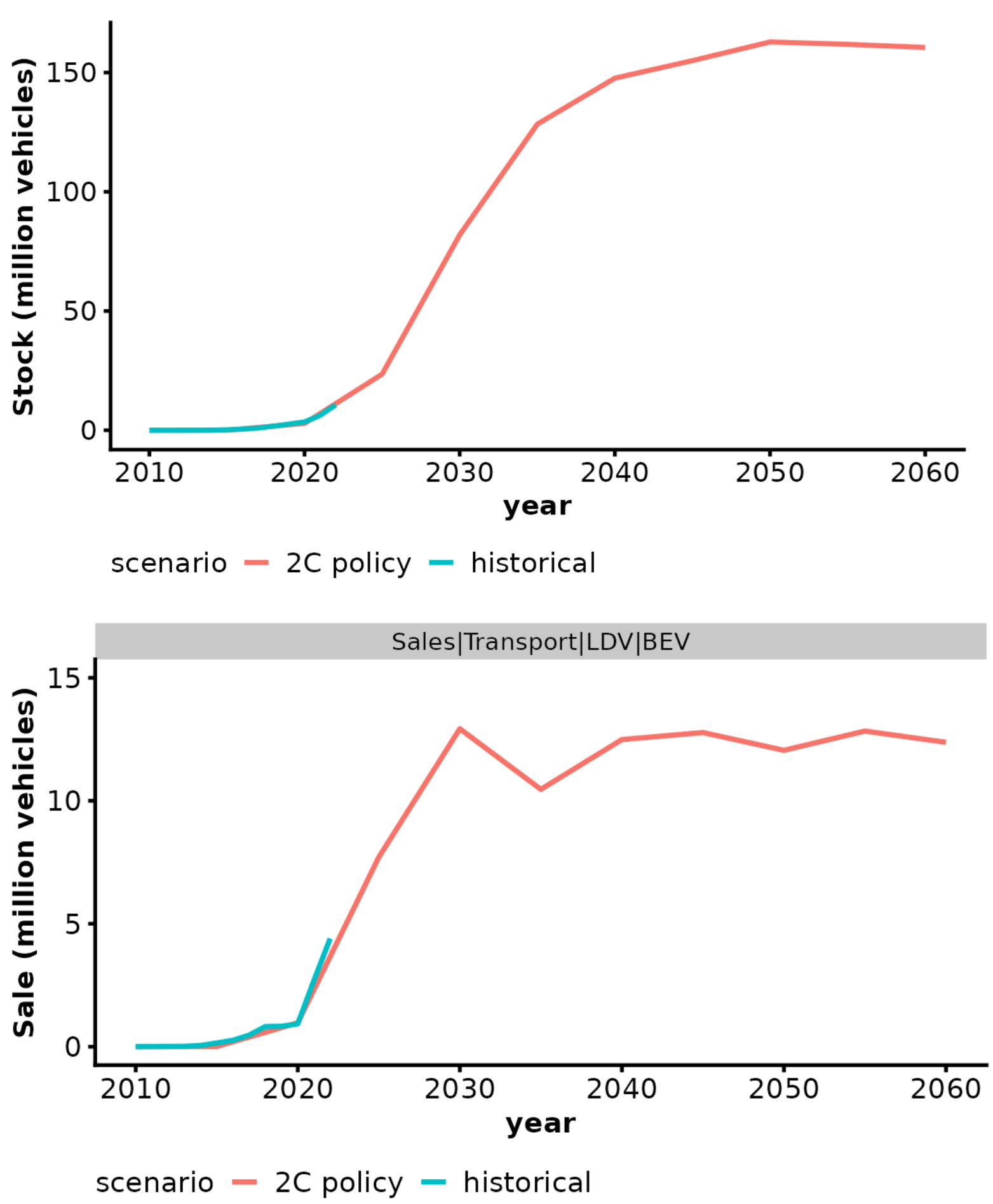

Figure S7: For light-duty BEV stock (top) and sales (bottom), the historical value (green) and scenario variables (red) are shown.

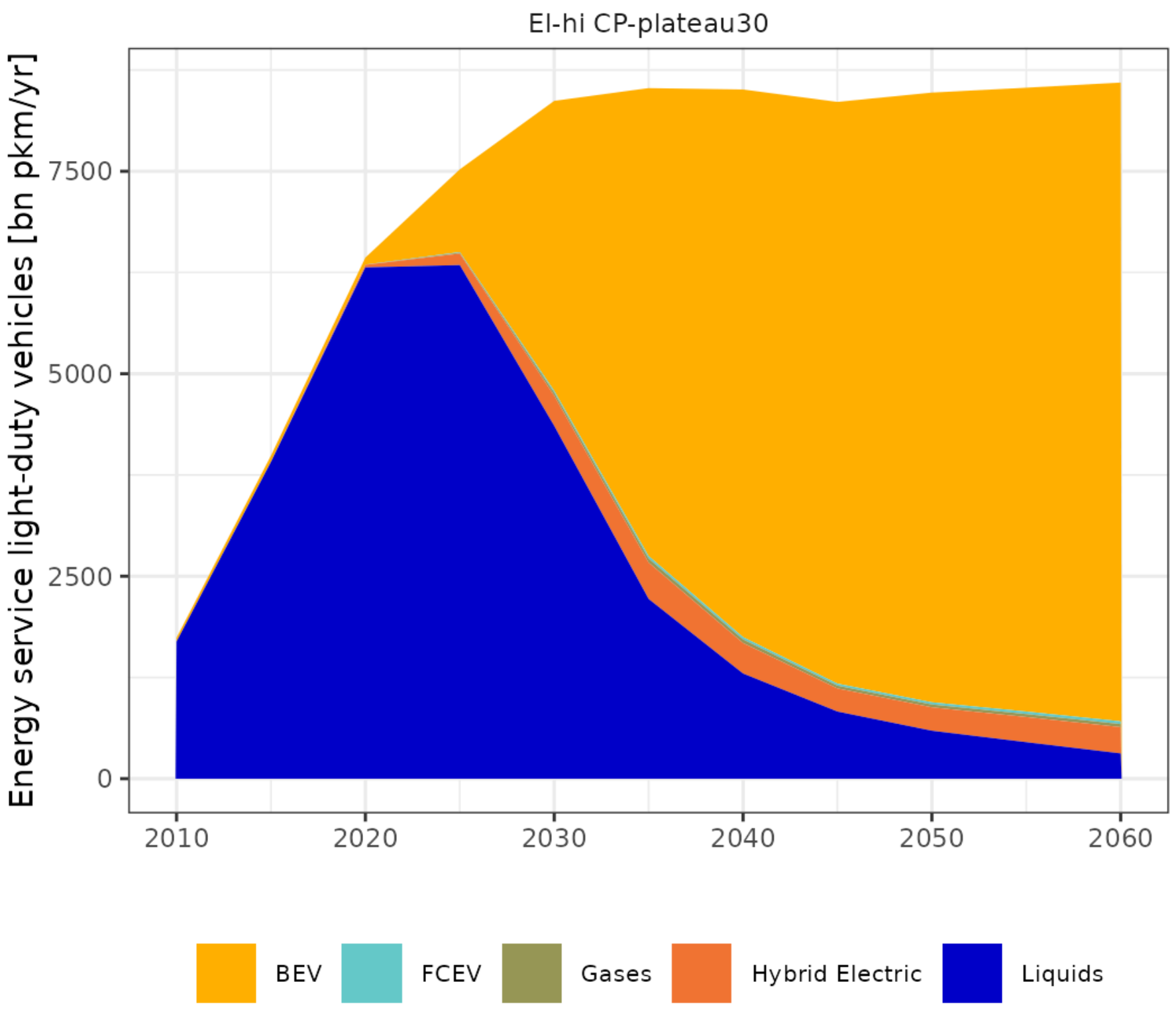

Figure S8: In a representative mitigation scenario ("El-hi CP-plateau30"), the energy service mix for light-duty vehicles shows that road passenger services are nearly fully electrified by 2050-2060, reaching around 91%.

As Figure S7 and S8 show, in the mitigation scenario for the transport sector, EV sales reach around 90% market share by 2030, with 8% of sales being hybrid electric vehicles. The ICE car fleet is nearly completely phased out by 2040. Thanks to strong, targeted policies, the electrification of transport remains robust, regardless of the pace of coal power phase-out. Our modeling scenario aligns with the latest data, which shows BEV sales reached 8.1 million in 2023.

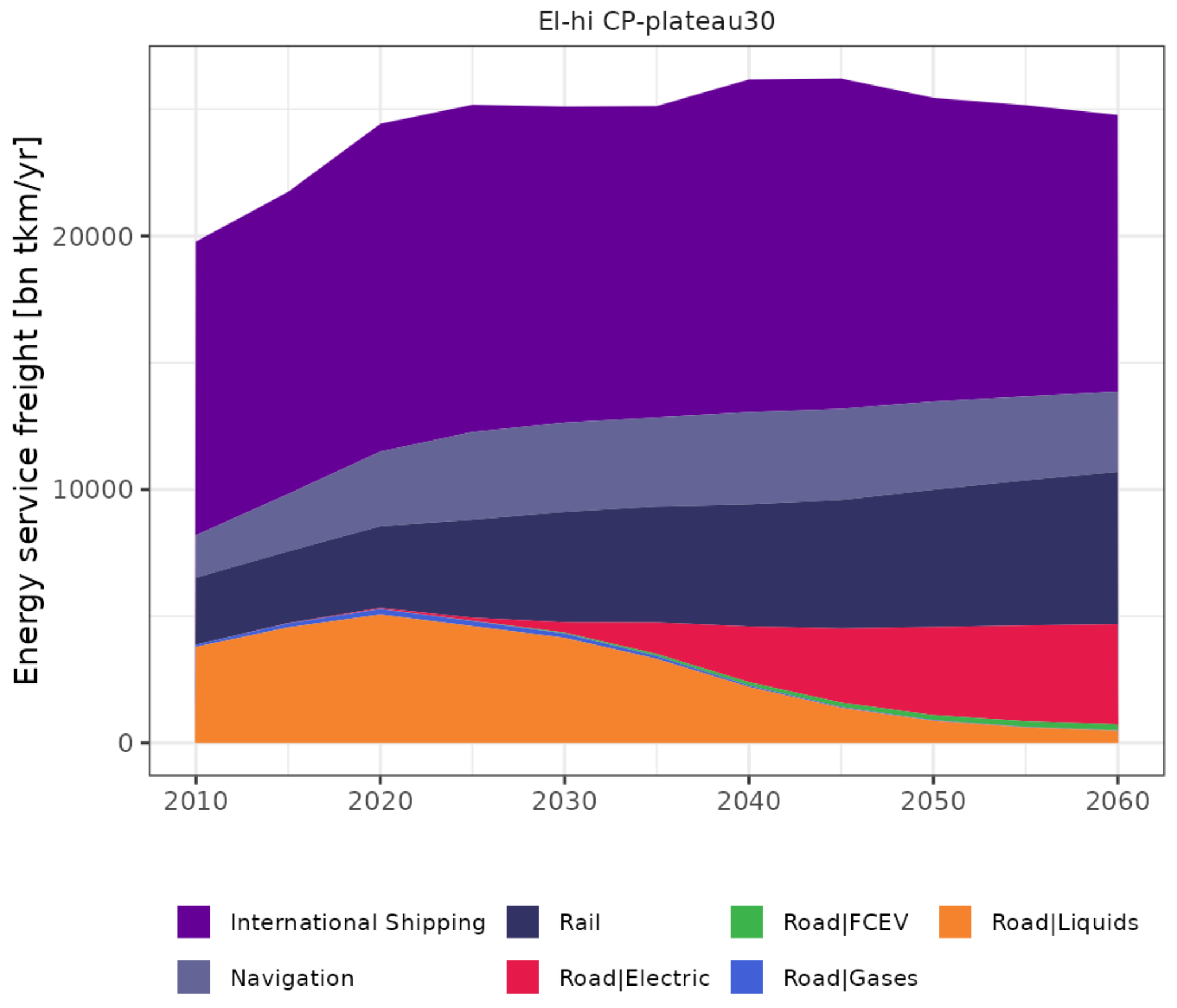

Figure S9: Energy service mix for freight vehicles and fleets in a representative mitigation scenario.

As shown in Figure S9, under the mitigation scenario, road freight is nearly fully electrified by 2060, reaching more than 80%. For simplicity, we assume that rail, inland navigation, and international shipping do not have the option to transition to end-use electrification.

8. Selected industry sector results

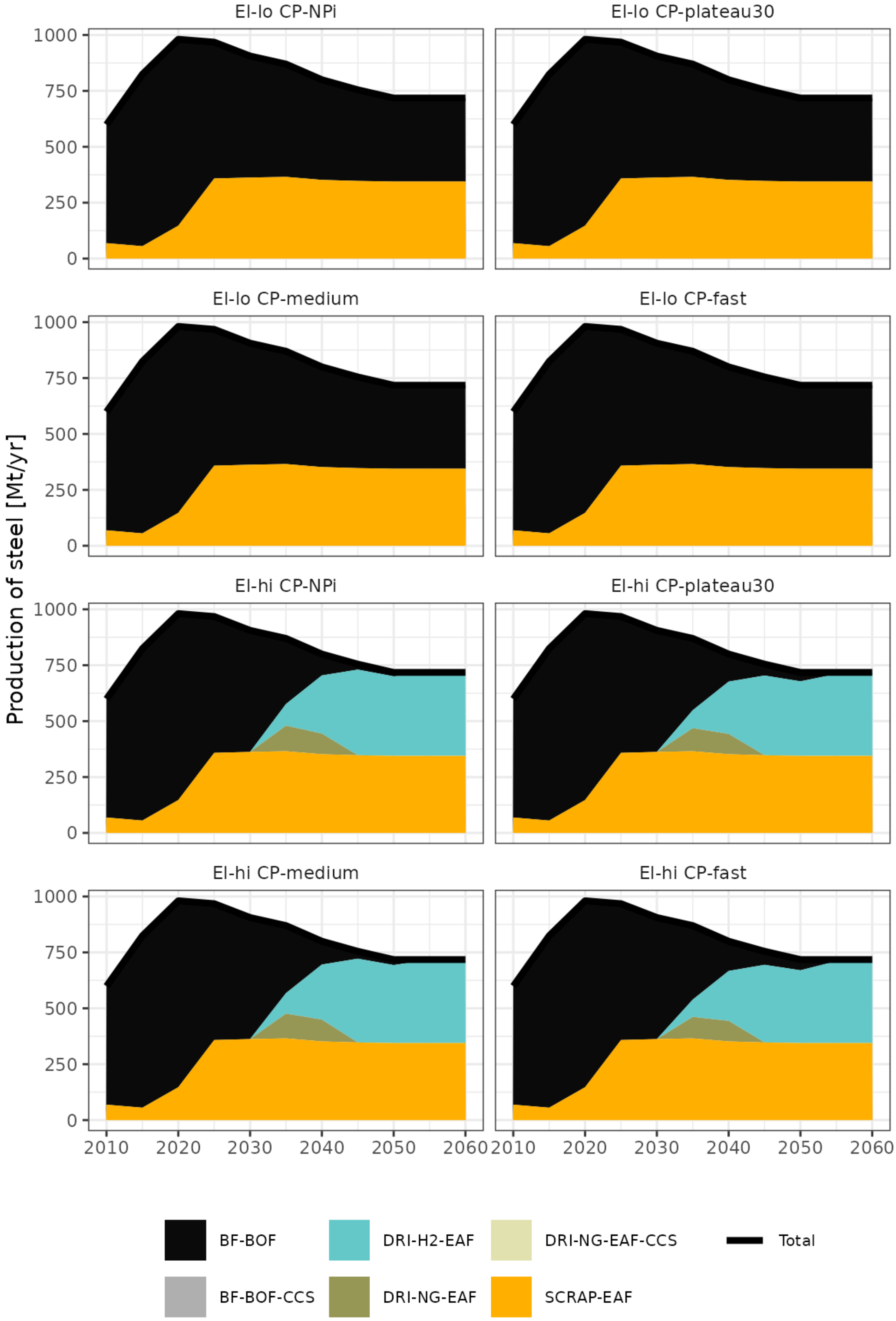

Figure S10: Steel production by process route.

As Figure S10 shows, in the transition of steel production, scrap-based secondary steel production plays a critical role in the near term (before 2025) in mitigation scenarios, driven by an increasing supply of scrap from end-of-life products. In contrast, the shift to hydrogen-based primary steel production takes much longer, as the large volume of primary steel production presents challenges for a rapid and complete transition to DRI-H2-EAF. This transition is further limited by the pace at which affordable green hydrogen production can be scaled up. A quick shift to DRI-H2-EAF would also incur the economic cost of prematurely retiring blast furnaces. During the phase-in period of DRI-H2-EAF, blast furnace emissions are partially reduced through the use of biomass solid (around 5%, not shown here), and for about ten years, natural gas replaces fossil coal due to the slow ramp-up of H2 electrolysis. Carbon capture does not play a major role in the scenarios shown, due to its high cost in blast furnace-based integrated steel plants and the competition among carbon-emitting sectors for limited geological carbon storage capacities. The resulting mix of steel processes aligns with Chinese steel scenarios found in published literature[6]. Across the mitigation scenarios, similar to investments in the transport sector, electricity price and mix are not key factors in determining investment decisions in the steel sector.

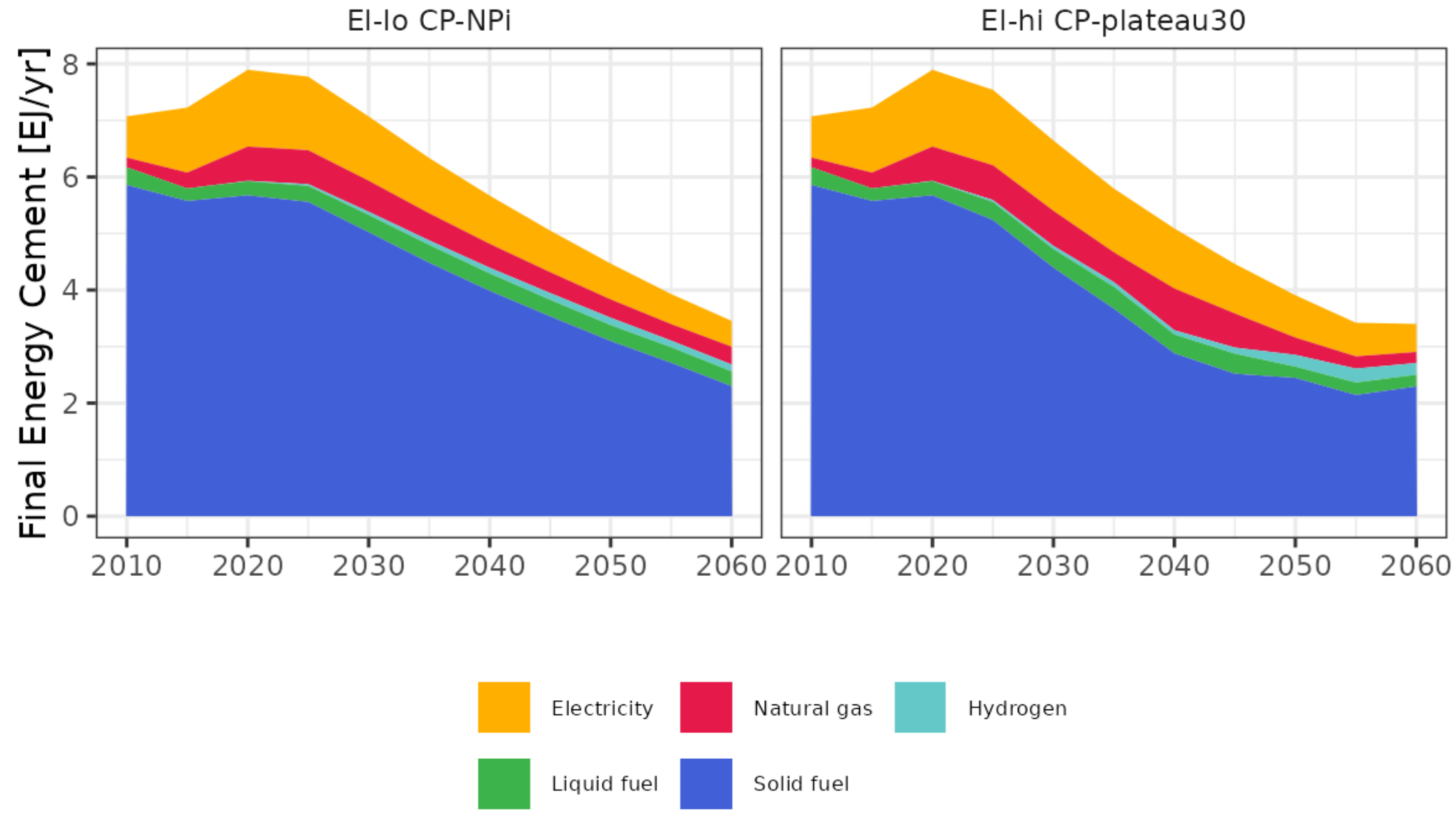

Figure S11: China's cement sector final energy mix in a representative reference scenario (left) compared to a representative climate policy scenario (right).

The annual demand/production of cement under both reference and mitigation scenarios is assumed to be the same, starting from today's production levels and reaching approximately 70% of that by 2050, or around 1500 Mt/yr. However, under the mitigation scenario, the overall final energy consumption decreases slightly. This reduction is driven by an endogenous increase in energy efficiency within the sector as a result of the climate constraint. The specific energy consumption for cement production decreases from around 3.1 GJ/t today to 2.2 GJ/t by 2060 under the climate constraint. In comparison, under the reference scenario, it decreases to 2.7 GJ/t by 2060.

Under climate constraints, the energy mix for cement shows a gradual shift towards electrification, with a rather small natural gas bridge in the 2040s, followed by a slight expansion of hydrogen in the mix after 2040, alongside electricity. Natural gas, coal, and oil are partially displaced by green hydrogen and electricity. The majority of solid fuels used are biogenic in origin, contributing to zero carbon emissions. Interestingly, due to the coal phase-out, high electrification in buildings, and the complete replacement of coal with hydrogen in the iron and steel sector, solid biomass used in the cement sector becomes the only remaining solid fuel in the entire Chinese economy.

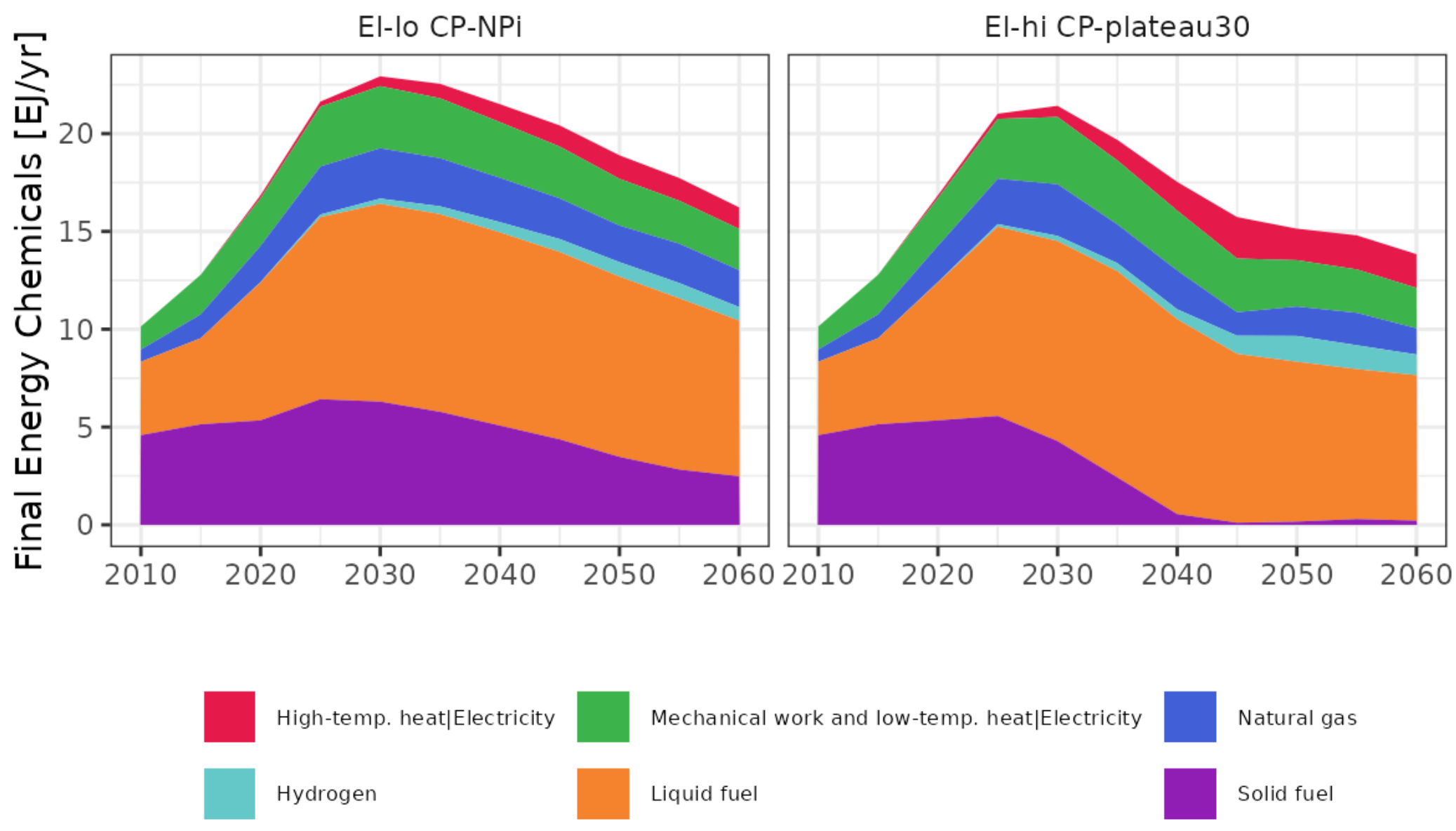

Figure S12: Final energy mix of China's chemicals sector in a representative reference scenario (left) and a representative climate policy scenario (right).

In the chemicals sector, under climate constraints, there is a gradual shift towards electrifying both high-temperature and low-temperature heat, as well as mechanical work. Natural gas, coal, and oil are partially displaced also by hydrogen. Despite annual output remaining the same, under climate constraint, the overall final energy consumption in the chemical sector decreases, driven by an endogenous increase in energy efficiency. Within liquid fuels, there is a slight increase in the share of biogenic hydrocarbon fuels, derived from bioenergy with carbon capture and storage, reaching around 10% (not shown here).

## 9. Scenarios with fixed carbon budget until 2060

The regional policy assumptions behind the scenarios we present in the paper consist only of net-zero $CO_2$ emission in 2060 in China, and no national emission budget targets are considered since there are currently no such domestic targets or pledges. However, due to lack of budget constraint, the expected trade-off between the timing of coal phase-out and the timing of electrification is not visible, nor is the climate policy reaction required to compensate for a later coal phase-out explicit, especially in terms of the expansion of negative emission technologies. In this section, in addition to the main scenario results we conduct further analysis under scenarios where the carbon budget between 2020 and 2060 is fixed.

To create this scenario, we keep the coal phase-out constraints in place, and in addition fix the carbon budget until net-zero to 210 $GtCO_2$. This budget corresponds to 18.2% of the global 2°C budget, comparable to the country's population which is 17.7% of the world total. Therefore it can be interpreted that the budget for China is aligned with equal per capita division of the remaining carbon budget.

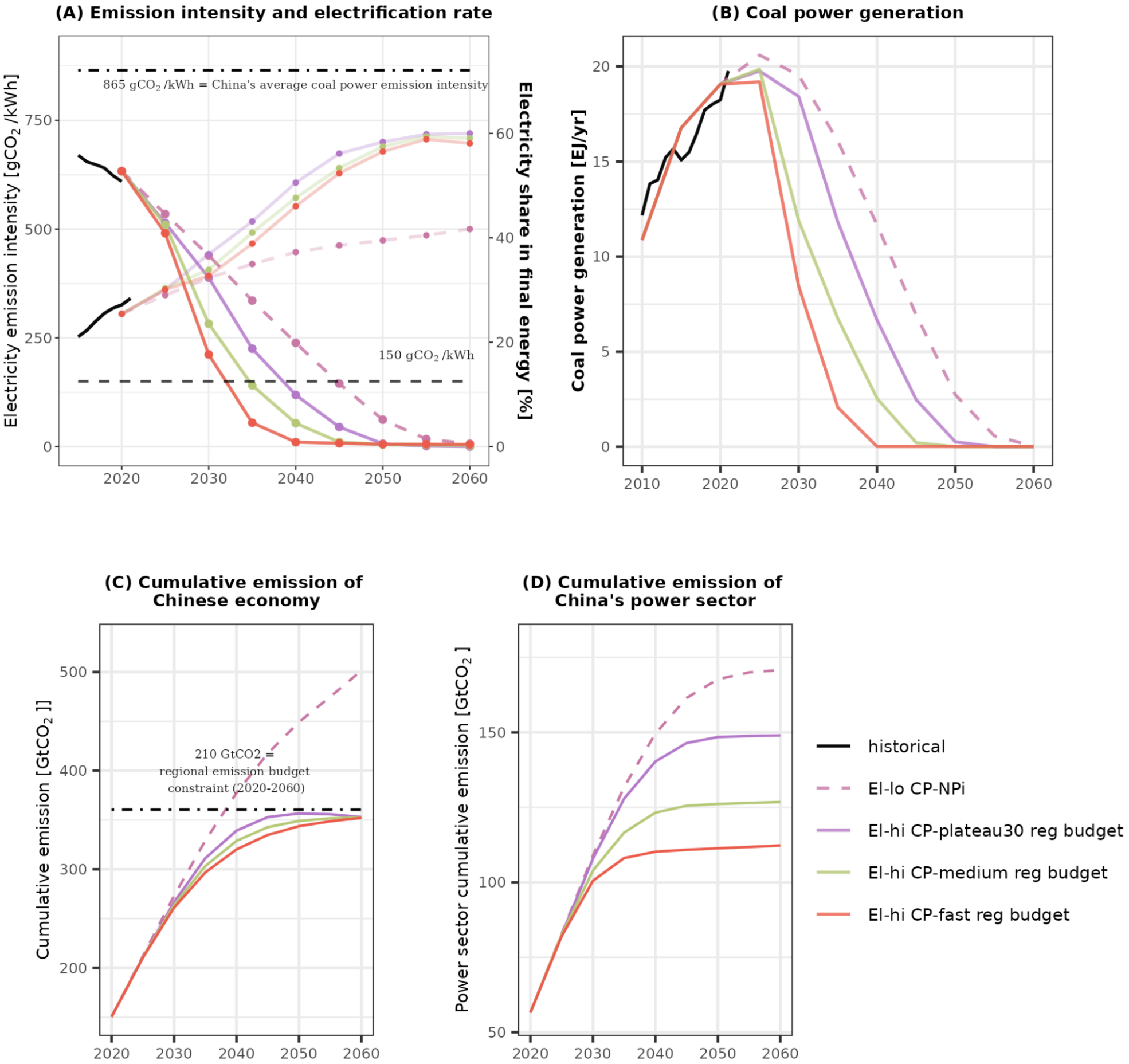

Figure S13: Cumulative emission results from a scenario with a fixed regional budget (emission target of 210 $GtCO_2$ from 2020 to 2060). All other constraints and the coal power trajectory assumptions are the same as the scenarios in the main text (as shown in Table 1). Here the reference scenario is “El-lo CP-NPi”, the other three correspond to the usual coal power phase-out settings, “El-hi CP-plateau30 reg(ional) budget”, “El-hi CP-medium reg budget”, and “El-hi CP-fast reg budget”, but now with a fixed emission budget in China .

The fixed-budget calculations result in different cumulative emissions from the power sector (Figure S13D), but similar fixed economy-wide cumulative emission in 2060 (Figure S13C). We observe a maximum difference of around 6% in electrification rate around 2040 to 2050 between “fast” and “plateau30” coal phase-out scenarios (Figure S13A). This is expected, since if coal phase-out happens faster, with a fixed budget, China will not need as ambitious an electrification target in the medium to short term, compared to a scenario where coal is phased out later.

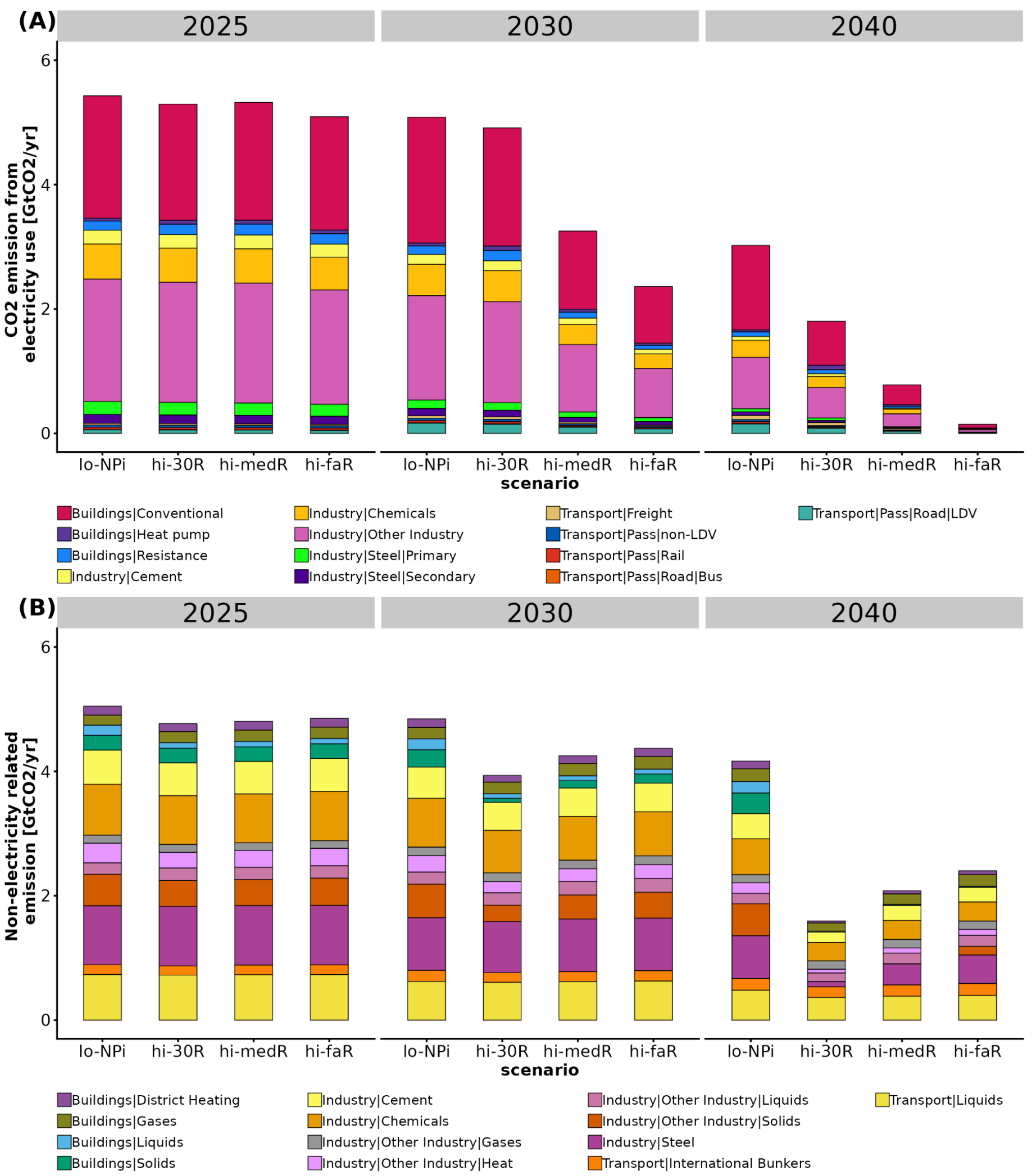

Figure S14: Annual $CO_2$ emissions from electrified vs. non-electrified buildings, industry, and transport subsectors, for the 4 scenarios: “El-lo CP-NPi” (“lo-NPi”), “El-hi CP-plateau30 reg budget” (“hi-30R”), “El-hi CP-medium reg budget” (“hi-medR”), and “El-hi CP-fast reg budget” (“hi-faR”).

As shown in Figure S14, varying levels of electrification lead to different annual emissions in the non-electricity sectors. This contrasts with the scenarios in the main text (Figure 5), where the electrification rate remains largely consistent across different coal phase-out speeds. In the regional fixed budget scenarios, direct electrification results in lower emissions by replacing coal, oil, and gas usage in non-electricity demand sectors (Figure S14B). If electrification were to raise emissions in China, a higher electrification rate would require a faster coal phase-out to stay within China’s fixed carbon budget. However, the opposite occurs: faster power sector decarbonization correlates with slower electrification under the fixed budget. Therefore, this finding aligns with our analysis in the main text, which shows that direct electrification with China’s current power mix does not lead to higher emissions.

The scenarios further show that, if coal phase-out is slower, electrification needs to be increased more, and a higher $CO_2$ price trajectory needs to be implemented due to the required negative emissions (Figure S15 and Figure S16).

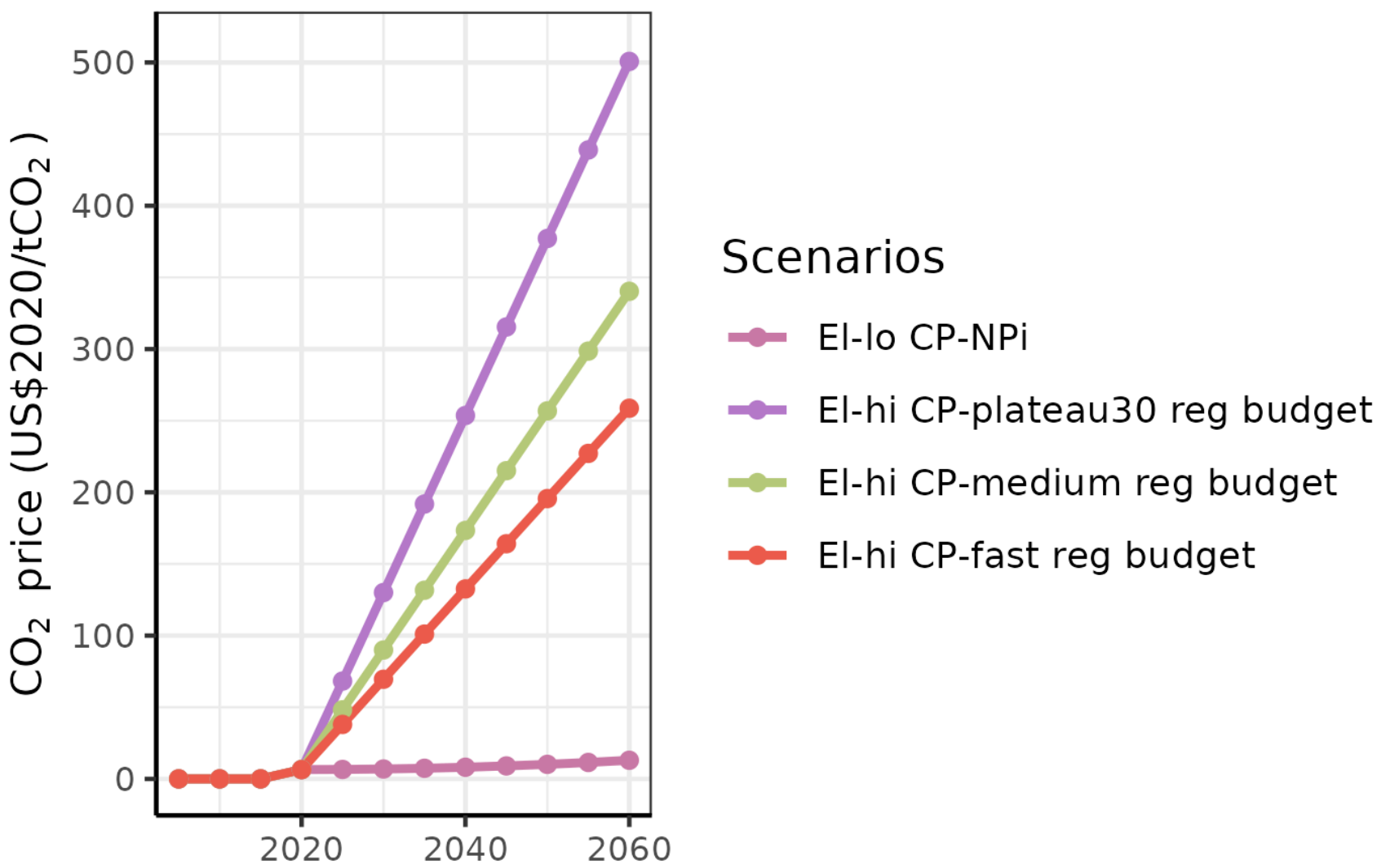

Figure S15: $CO_2$ price trajectory for China under a fixed $CO_2$ emission target of 210 $GtCO_2$ from 2020 to 2060.

As shown in Figure S15, with a fixed regional budget, when coal phase-out happens faster, a lower endogenous $CO_2$ price trajectory is obtained. When coal phase-out happens slower, $CO_2$ price is higher (by > 200 $/$tCO_2$ in 2060 compared to faster coal phase-out). Because in REMIND, coal power generation is almost never competitive with renewables, coal is phased out under baseline economic fundamental assumptions. So the faster phase-out actually incurs a lower $CO_2$ price in REMIND, because $CO_2$ price is not needed to phase out coal power in the model due to the cost-competitiveness of renewables, but is needed to phase in direct and indirect electrification, as well as negative emission. Hence as power sector cumulative emission gets bigger, the non power sector targets become more stringent, resulting in higher $CO_2$ price.

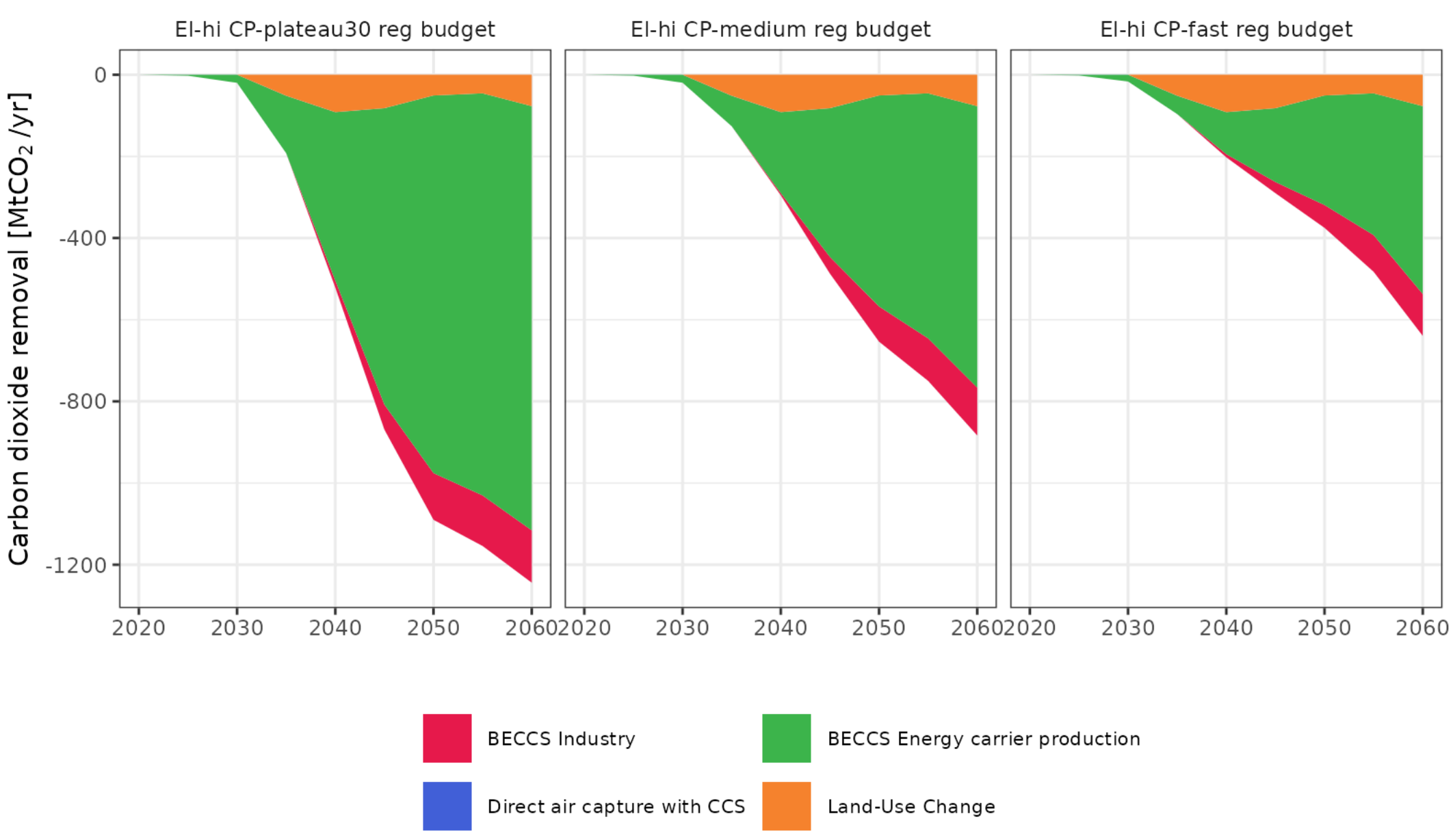

Figure S16: Negative emission for fixed regional budget scenarios under various coal generation phase-out pathways.

As shown in Figure S16, under a fixed $CO_2$ emission target, when coal phase-out happens slower, negative emission technologies need to be ramped up faster after 2030 to reach the same target than ambitious coal phase-out scenarios. Given that this budget corresponds roughly to equal per capita remaining budget allocation, a ten-year delay in the coal phase-out corresponds to an almost doubling of negative $CO_2$ emission in 2060, from around 600 $MtCO_2$ to 1.2 $GtCO_2$ in 2060. Most of the negative emissions are supplied via bioenergy with carbon capture and storage.

10. China's cumulative emission budget in the context of historical emission and global structural inequality

In this study, the way of accounting for cumulative emissions only starting from 2023 and comparing it to the remaining carbon budget until the end of the century obviously ignores the issues with historical emissions before 2023, where China compared to Europe and the U.S. has currently a relatively smaller share, even as on a population aggregate basis[7]. The cost-optimizing mitigation scenarios from IAMs typically exclude considerations of international climate equity such as differentiated responsibilities and respective capabilities, as well as under the global structural inequality highlighted by "unequal ecological exchange" where China links production and consumption of different parts of the world as an export and manufacturing powerhouse[8–11]. In addition, with the official pledge of peaking before 2030, the country would be surpassing other industrialized economies in achieving per capita emission peaking at a low level of per capita emission and a low level of development indicated by GDP per capita. The Chinese per capita annual emission would be around 7-8 $tCO_2$ when it peaks, while other economies during their peaks emitted 7-15 $tCO_2$ (Germany 14 $tCO_2$, U.S. 20 $tCO_2$ and Japan 10 $tCO_2$)[12].

11. Policies to safeguard the power-sector transitions

Due to the high emission intensity nature of the Chinese power grid and the electrification revolution which is on the verge of take-off, China's power sector transition needs a robust time plan. Besides continuing to push for existing policies such as power market reform, increasing inter-provincial power trading, ETS, and non-fossil fuel targets[13,14], risk-management needs to be an essential component of power sector planning. Both supply-side and demand-side policies should be devised to avoid unnecessary bottlenecks, prevent market disturbances and shocks.

On the production side, policies to prevent material and energy shortages could be designed to safeguard the transition. First, as all the world's industrialized economies start to ramp up their mitigation efforts, resource constraints will almost inevitably happen. Therefore access to raw materials could be planned ahead of time in order to minimize supply chain bottlenecks in photovoltaic (PV) modules, wind turbines and batteries. A publicly owned strategic reserve of necessary mineral resources used for the production of mitigation technologies could be created to dampen price increase in times of shortage. Second, with the existing power system, energy shortages should be avoided. Existing and newly built power systems, especially transmission and combustion generation need to be climate-proof to prevent outages from fire and droughts. Hydroelectric power in China is already adversely impacted by climate change, as demonstrated in the case of highly hydro-dependent provinces Yunnan and Sichuan in 2022, when local rainfall has dropped by about two-thirds from the same period, which led to energy shortages as industrial and residential demand surged. Because many green industries are in these locations to take advantage of the low-carbon hydro power, this energy shortage eventually led to polysilicon shortages and price spikes that impacted and disturbed the PV module supply chain. Besides the policy responses such as from the National Energy Administration (NEA) in the aftermath of the shortage: speeding up nuclear, transmission and hydro power station constructions, as well as increasing grid flexibility[15], policymakers should think more strategically about the locations of the crucial parts of the supply chain for manufacturing renewable components such as battery and silicon which are power intensive. To avoid adverse impact of these productions, the locations of these industries are advised to be diversified to hedge against the risk of energy shortage. This ensures the build-out of the green infrastructure is not delayed by the risk of the existing one.

On the demand-side, better weather forecasts and climate impact projections should be done to better predict unexpected heat waves and demand surges, such as for air conditioning. As more and more battery-equipped vehicles are on the road, policies around automated demand-side managements such as smart-charging and vehicle-to-grid should be given priority for deployment, and these technologies should be given as mandates to EV manufacturers and implemented at scale to work as part of the storage asset of the power grid[16].

12. Capital cost of key power sector technologies in scenarios

Table S3: Capital costs of key power generation technologies, including Solar PV, onshore Wind, and offshore Wind. Due to space limitations, endogenous cost changes are only shown for a representative set of scenarios. Costs, other than for Solar PV, have been rounded to the nearest tenth.

| **Overnight capital costs (US$2020/kW)** | **Scenario** | **2020** | **2030** | **2040** | **2050** | **2060** |
|---|---|---|---|---|---|---|
| **Pulverized coal plant** | All | 615 | 615 | 615 | 615 | 615 |
| **Nuclear plant** | All | 4230 | 4150 | 3770 | 3770 | 3770 |
| **Solar PV** | El-lo CP-NPi | 867 | 521 | 440 | 425 | 391 |
| | El-hi CP-plateau30 | 867 | 482 | 402 | 395 | 367 |
| | El-hi CP-medium | 867 | 472 | 400 | 394 | 367 |
| | El-hi CP-fast | 867 | 467 | 399 | 393 | 366 |
| **Wind Onshore** | El-lo CP-NPi | 1280 | 1190 | 1150 | 1150 | 1070 |
| | El-hi CP-plateau30 | 1280 | 1140 | 1080 | 1080 | 1010 |
| | El-hi CP-medium | 1280 | 1120 | 1070 | 1070 | 1010 |
| | El-hi CP-fast | 1280 | 1120 | 1070 | 1070 | 1010 |
| **Wind Offshore** | El-lo CP-NPi | 5190 | 3640 | 2570 | 1900 | 1790 |
| | El-hi CP-plateau30 | 5190 | 3430 | 2410 | 1830 | 1740 |

| | El-hi CP-medium | 5190 | 3410 | 2410 | 1830 | 1740 |
|---|---|---|---|---|---|---|
| | El-hi CP-fast | 5190 | 3410 | 2410 | 1830 | 1740 |

The costs of green technologies vary under different coal phase-out scenarios due to endogenous learning. Under the REMIND assumptions, regional costs eventually converge to a single cost floor. As a result, technical costs in developing countries with currently low costs may rise slightly due to factors such as increasing wages and stricter environmental and quality regulations. For onshore wind, China is already close to the global cost floor, allowing it to converge to this level more quickly. In contrast, Europe and the US will experience a slower convergence to the cost floor.

13. Electrification rate compared to AR6 1.5°C scenarios

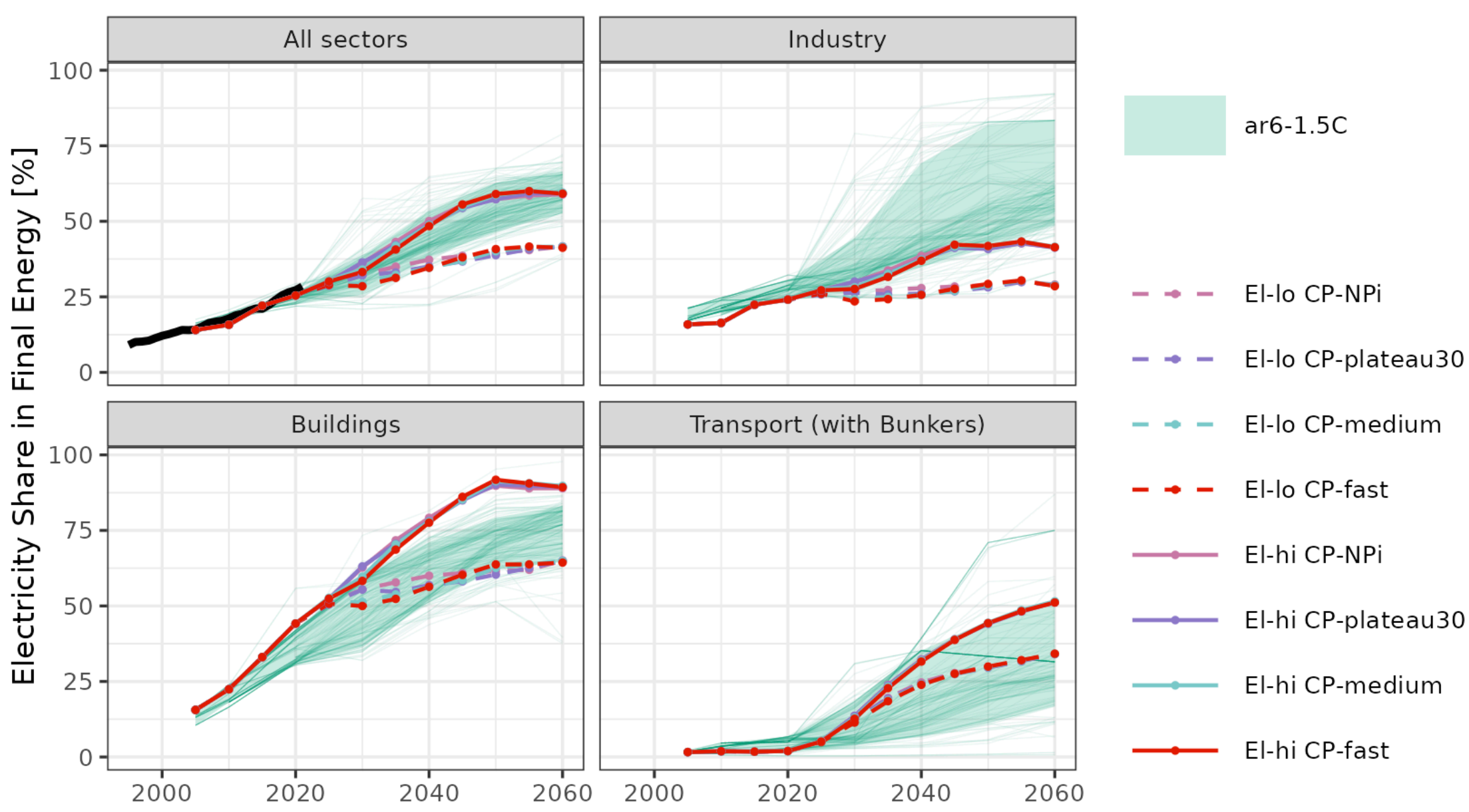

Figure S17: Economy-wide electrification rate and the electrification rates of the three end-use sectors under the scenarios in the paper, compared to IPCC AR6 scenario data for the 1.5°C scenarios. The thin lines indicate all AR6 scenarios for 1.5°C, and the darker area indicates the 10-90% quantiles. For the R10 CHINA+ regions, the regional scenarios C1 and C2 are selected for the global 1.5°C scenario. Related to Figure 3.

Compared to the AR6 1.5°C scenarios our scenarios' overall electrification rate is somewhat higher (by around 8%) in the medium term in the 2040s. However, the transport electrification rate is high (equivalent to the top 90% quantile of AR6 scenarios) due to the high shares of electrification of trucks (above 90%), not shown, industry is low (10% quantile), and building is very high (above 90% quantile) due to the wide application of heat pumps to substitute wood or coal.

The IAM results from IPCC AR6 database shown here are publically available[17], and contain metadata including the assumptions of their scenarios. The main assumptions are parameters such as technology costs, biomass availability and cost, total potential and cost of negative emission technologies. For example, if large-scale negative emission technologies such as bioenergy with carbon capture and storage (BECCS) is available, then there could be more residual emissions from hard to abate sectors (which are some of the hardest to electrify). This would reduce the electrification rate.

14. Comparison of carbon emission intensity between European countries and China's provinces

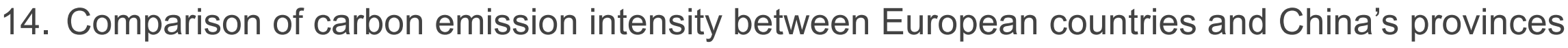

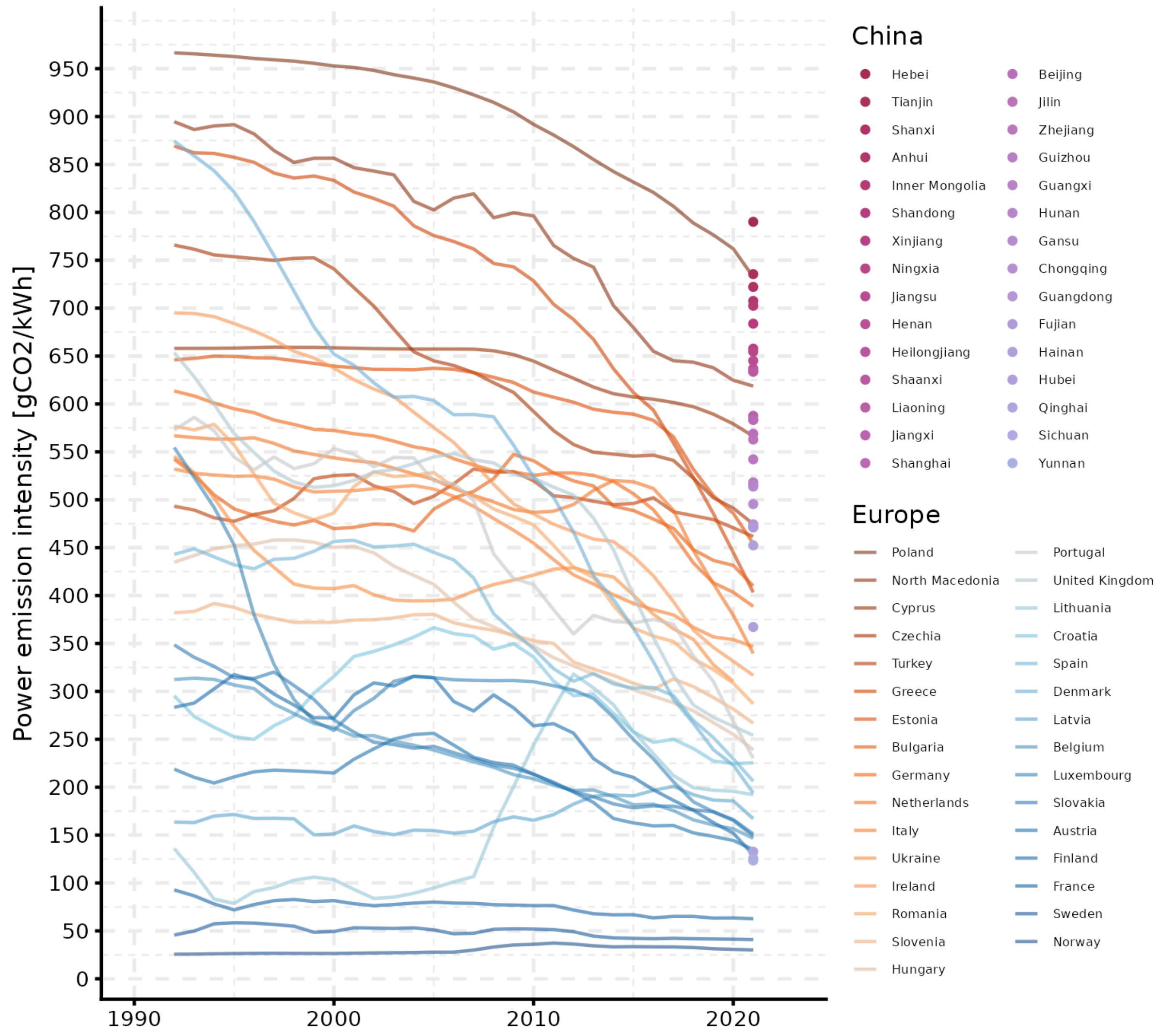

Figure S18: Carbon emission intensity of electricity for European countries and Chinese provinces. For better visualization, the data has been smoothened from the data source. Data sources: Ember (2024) and Energy Institute - Statistical Review of World Energy (2024) – with major processing by Our World In Data[18,19].

As shown in Figure S18, for many European countries, in the last ten years, there has been an acceleration of decrease of emission intensity of electricity from roughly 500 to 850 gCO$_2$/kWh to today's 250 to 500 gCO$_2$/kWh. In contrast, Chinese provinces' intensities of electricity remain high, roughly on par with the European countries between early 1990s and 2000. Most European countries have decreased their intensity by a significant amount over the last 30 years. Another noticeable difference is that while in 1995, roughly half of the European countries already were below 500 gCO$_2$/kWh, due to higher hydro and nuclear shares, only nine Chinese provinces are below 500 gCO$_2$/kWh today.

15. Cumulative emissions of China’s sectors and their climate impact

Table S4: Cumulative emission and peak temperature impact from supply and demand sectors under various electrification and coal phase-out scenarios.

<table>
<tr><th colspan="2" rowspan="3">Scenario</th><th colspan="11">$CO_2$ emission (2023-2060) [Gt] | temperature contribution [°C]</th></tr>
<tr><th colspan="2" rowspan="2">Power sector [Gt] | [°C]</th><th colspan="6">Demand sector (energy)</th><th rowspan="2">Sum of non-power energy sectors [°C]</th><th colspan="2" rowspan="2">Economy [Gt] | [°C]</th></tr>
<tr><th colspan="2">Buildings [Gt] | [°C]</th><th colspan="2">Transport [Gt] | [°C]</th><th colspan="2">Industry [Gt] | [°C]</th></tr>
<tr><td rowspan="4"><b>Reference / No climate constraint</b></td><td><b><i>El-lo CP-NPi</i></b>: 2055 coal phase-out, low elec.</td><td>99</td><td>0.045</td><td>25</td><td>0.011</td><td>20</td><td>0.009</td><td>91</td><td>0.041</td><td>0.061</td><td>315</td><td>0.142</td></tr>
<tr><td><b><i>El-lo CP-plateau30</i></b>: 2050 coal phase-out, low elec.</td><td>78</td><td>0.035</td><td>27</td><td>0.012</td><td>19</td><td>0.009</td><td>96</td><td>0.043</td><td>0.064</td><td>302</td><td>0.136</td></tr>
<tr><td><b><i>El-lo CP-med</i></b>: 2045 coal phase-out, low elec.</td><td>55</td><td>0.025</td><td>27</td><td>0.012</td><td>19</td><td>0.009</td><td>97</td><td>0.044</td><td>0.065</td><td>281</td><td>0.127</td></tr>
<tr><td><b><i>El-lo CP-fast:</i></b> 2040 coal phase-out, low elec.</td><td>39</td><td>0.018</td><td>27</td><td>0.012</td><td>19</td><td>0.009</td><td>97</td><td>0.044</td><td>0.065</td><td>267</td><td>0.120</td></tr>
<tr><td rowspan="3"><b>Mitigation/ Net-zero</b></td><td><b><i>El-hi CP-NPi</i></b>: 2055 coal phase-out, high elec.</td><td>99</td><td>0.044</td><td>7</td><td>0.003</td><td>15</td><td>0.007</td><td>44</td><td>0.020</td><td>0.030</td><td>199</td><td>0.090</td></tr>
<tr><td><b><i>El-hi CP-plateau30</i></b>: 2050 coal phase-out, high elec.</td><td>78</td><td>0.035</td><td>7</td><td>0.003</td><td>15</td><td>0.007</td><td>44</td><td>0.020</td><td>0.030</td><td>178</td><td>0.080</td></tr>
<tr><td><b><i>El-hi CP-med</i></b>: 2045 coal phase-out, high elec.</td><td>55</td><td>0.025</td><td>7</td><td>0.003</td><td>16</td><td>0.007</td><td>45</td><td>0.020</td><td>0.030</td><td>158</td><td>0.071</td></tr>
</table>

| | | | | | | | | | | | | |
|---|---|---|---|---|---|---|---|---|---|---|---|---|
| | ***El-hi CP-fast:*** 2040 coal phase-out, high elec. | 38 | 0.017 | 7 | 0.003 | 15 | 0.007 | 45 | 0.020 | 0.030 | 143 | 0.064 |

16. Illustrative diagrams of the REMIND module

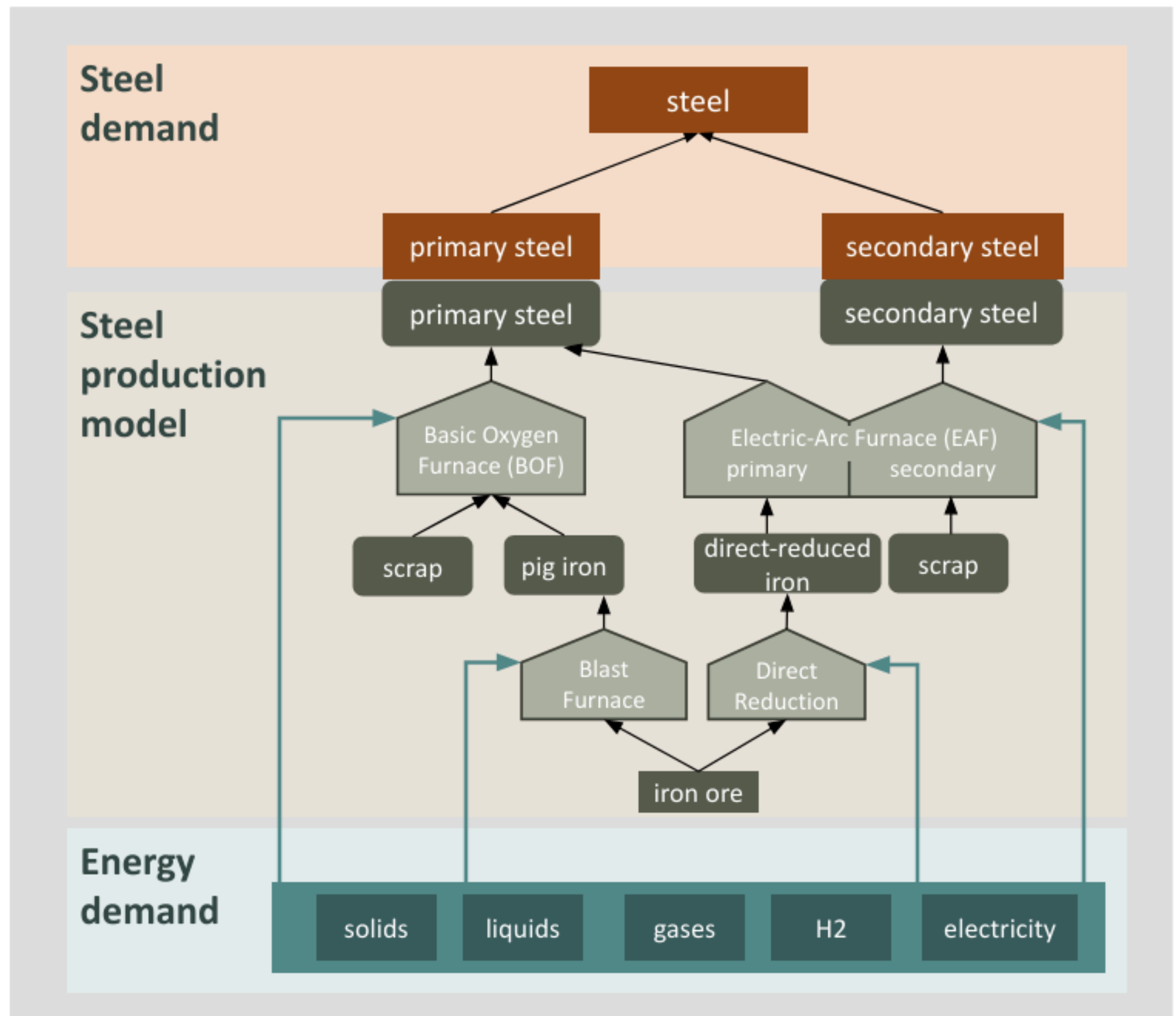

Figure S19: Illustration of the process-based industry module steel submodule structure of REMIND v3.3.0.dev132.

In the new formulation of the steel sector in REMIND v3.3.0.dev132. industrial processes are explicitly represented, in contrast to the old CES based formulation. Raw material (i.e. iron ore) and their cost is considered as well as the capital investment of the production capacities (e.g. BOF, EAF, blast furnace or direct reduction). This provides a more bottom-up approach to more accurately calculate steel demand and steel production.

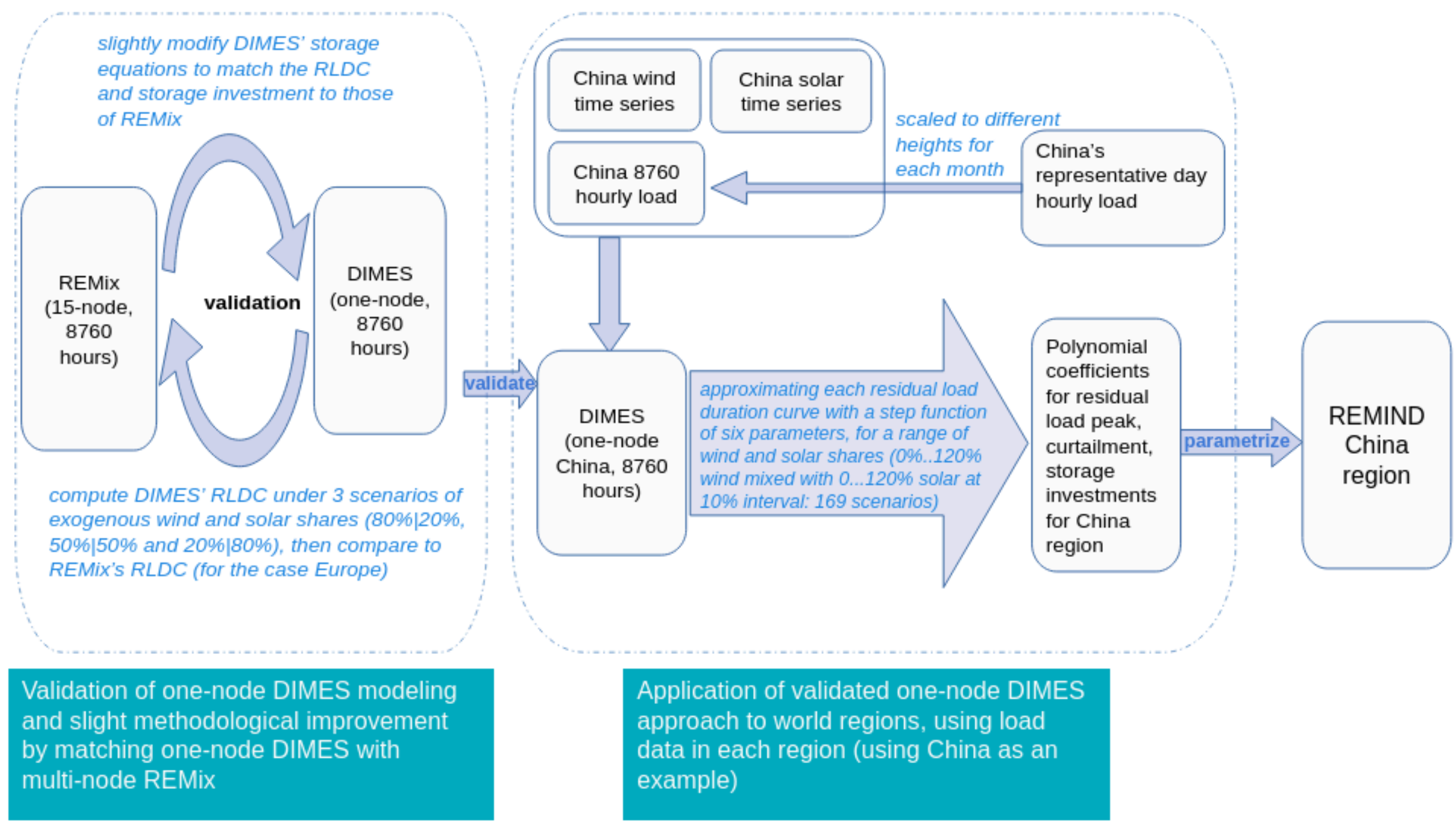

Figure S20: Flow diagram describing the process of the reparametrization of REMIND aggregate regional representation of wind and solar, as carried out in the Ueckert et al. 2017 study[20].

The parametrization of the REMIND power sector in China used the hourly single-node model DIMES, based on China's hourly load and renewable weather data to find the parametric representation of storage, curtailment and residual load peak for the Chinese region, which feeds into the REMIND equations. Since REMix's geographical coverage is limited to Europe, the model is only used in the validation step for one-node model DIMES. It was not used to parameterize other world regions.

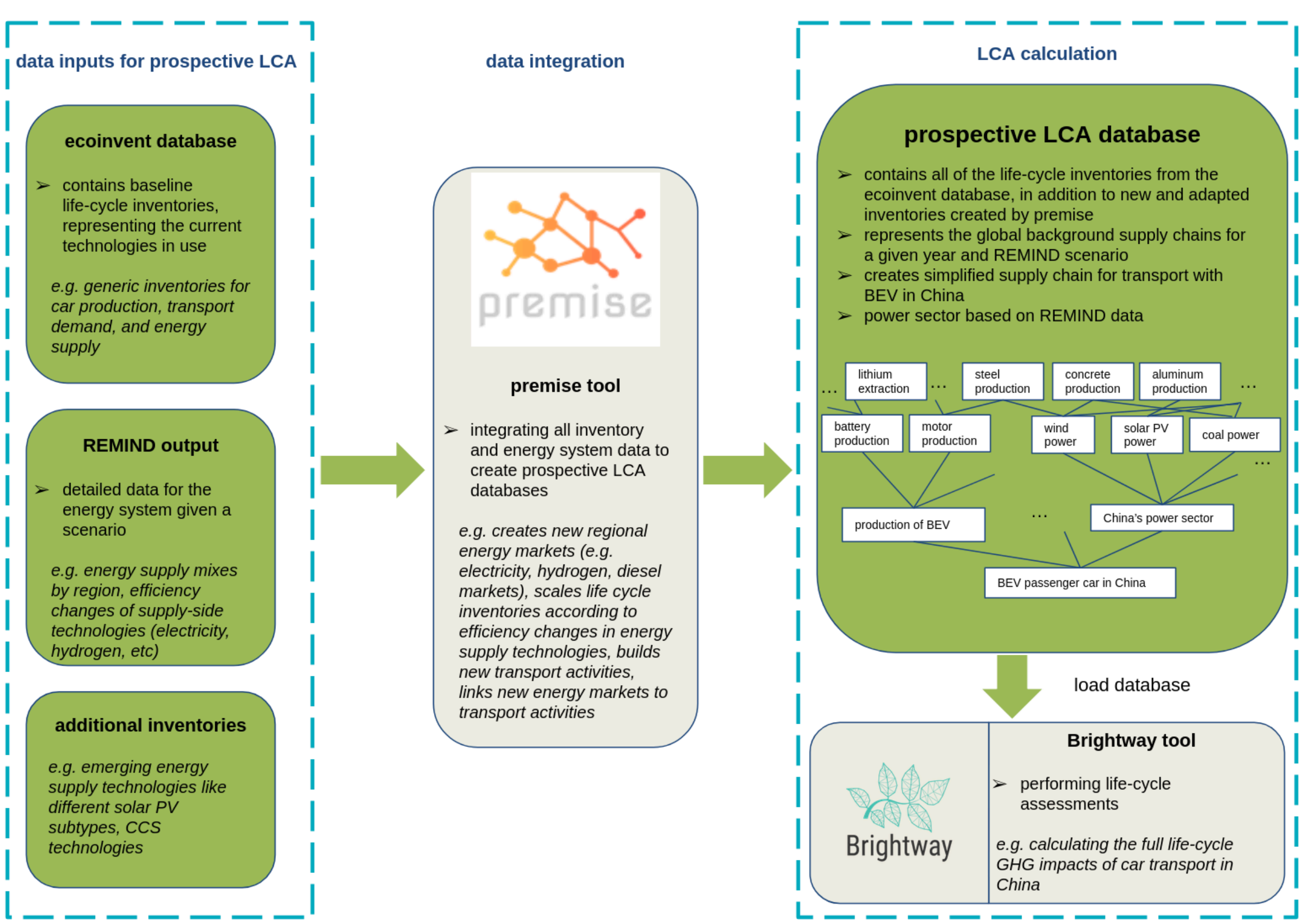

Figure S21: Workflow for integrating IAM and the prospective LCA.

In the REMIND LCA module, the source database ecoinvent is enhanced using data from the REMIND model and additional life cycle inventories. The necessary data integration and creation of new, prospective databases is done by the software tool *premise.* The new databases are directly exported to be used with the LCA software *Brightway,* where the assessments (functional units, impact method) are set up and calculated.

Supplemental references